\documentclass[final,5p,times,twocolumn,colorlinks,citecolor=DarkGreen,linkcolor=DarkRed,urlcolor=DarkBlue,pdfencoding=auto, psdextra]{elsarticle}

\usepackage{lipsum}
\usepackage{mathtools}
\usepackage{cuted}\usepackage{graphicx}
\usepackage{amssymb}
\usepackage{amsmath}
\usepackage[svgnames]{xcolor}
\usepackage{mathtools,slashed}
\usepackage[retainorgcmds]{IEEEtrantools}
\usepackage{physics}
\usepackage[compat=1.1.0]{tikz-feynhand}
\usepackage{tikz}
\usepackage{epstopdf}
\usepackage[utf8]{inputenc}
\usepackage{url}
\usepackage[colorlinks,citecolor=DarkG reen,linkcolor=DarkRed,urlcolor=DarkBlue]{hyperref}

\usepackage{newtxtext}
\usepackage{newtxmath}
\usepackage{tabularx}
\usepackage{booktabs}
\usepackage{bookmark}
\usepackage{subfig}
\usepackage{makecell}

\usepackage{float}

\setcitestyle{square}

\journal{Physics Letters B}

\begin{document}

\begin{frontmatter}

\title{Direct measurement of the $^{103}$Rh(n,$\gamma$) and $^{103}$Rh($\gamma$,n) cross section up to stellar  temperatures at the CSNS Back-n and SSRF SLEGS}
\author[label1,label2]{Hao Liang}
\author[label1,label3,label4]{Zhen-dong An\corref{cor1}}
\ead{anzhendong@impcas.ac.cn}
\cortext[cor1]{Corresponding authors}
\author[label5,label6]{Wei Jiang}
\author[label7,label8]{Zi-rui Hao}
\author[label2]{Chen-chen Guo\corref{cor1}}
\ead{guochch7@mail.sysu.edu.cn}
\author[label3,label4]{Yu-gang Ma\corref{cor1}}
\ead{mayugang@fudan.edu.cn}
\author[label9]{Jie Ren}
\author[label9]{Xi-chao Ruan}
\author[label10]{Jing-yu Tang}
\author[label5,label6]{Rui-rui Fan}
\author[label7,label8]{Gong-tao Fan}
\author[label7,label8]{Hong-wei Wang}
\author[label7,label8]{Wen-qing Shen}
\author[label1,label13]{Yu-bing Li}
\author[label1,label2]{Jun-heng Hu}
\author[label1,label11]{Di Sun}
\author[label1,label2]{Ting Liu}
\author[label1,label2]{Zi-jun Liu}
\author[label1,label12]{Yi Sui}

\address[label1]{Institute of Modern Physics, Chinese Academy of Sciences, Lanzhou 730000, China}
\address[label2]{Sino-French Institute of Nuclear Engineering and Technology, Sun Yat-sen University, Zhuhai 519082, China}
\address[label3]{Key Laboratory of Nuclear Physics and Ion-beam Application (MOE), Institute of Modern Physics, Department of Nuclear Science and Technology, Fudan University, Shanghai 200433, China}
\address[label4]{Shanghai Research Center for Theoretical Nuclear Physics, NSFC and Fudan University, Shanghai 200438, China}
\address[label5]{Institute of High Energy Physics, Chinese Academy of Sciences, Beijing 100049, China}
\address[label6]{Spallation Neutron Source Science Center, Dongguan 523803, China}
\address[label7]{Shanghai Advanced Research Institute, Chinese Academy of Sciences, Shanghai 201210, China}
\address[label9]{Shanghai Institute of Applied Physics, Chinese Academy of Sciences, Shanghai 201800, China}
\address[label9]{Key Laboratory of Nuclear Data,China Institute of Atomic Energy, Beijing 102413, China}
\address[label10]{School of Nuclear Science and Technology, University of Science and Technology of China, Hefei 230027, China}
\address[label11]{College of Physical Science and Technology, Shenyang Normal University, Shenyang, 110034, China}
\address[label12]{School of Nuclear Science and Technology, University of South China, Hengyang 421001, China}
\address[label13]{Guangxi Key Laboratory of Nuclear Physics and Technology, College of Physics and Technology, Guangxi Normal University, Guilin 541004, China}

\begin{abstract}

The cross sections of $^{103}$Rh(n,$\gamma$) and $^{103}$Rh($\gamma$,n) play a crucial role in the stellar nucleosynthesis, rhodium-based self-powered neutron detectors, and nuclear medicine. The cross sections of $^{103}$Rh(n,$\gamma$) was measured by the time-of-flight(TOF) method from 1 eV to 1000 keV at the Back-n facility of the Chinese Spallation Neutron Source. In the resolved resonance region, the data reported multiple new resonance structures for the first time. And some discrepancies were observed, offering valuable insights into the differences between the evaluated libraries. Maxwellian-averaged cross sections (MACSs) were calculated within the temperature range of the $s$ process nucleosynthesis model, based on the averaged cross sections in the unresolved resonance region. Meanwhile the cross sections of $^{103}$Rh($\gamma$,n) within the range of $p$ process nucleosynthesis were measured using laser Compton scattering (LCS) $\gamma$ rays and a new neutron flat efficiency detector (FED) array at the Shanghai Laser Electron Gamma Source (SLEGS), Shanghai
Synchrotron Radiation Facility (SSRF). Using an unfolding iteration method, $^{103}$Rh($\gamma$,n) data were obtained with uncertainty less than 5\%, and the inconsistencies between the available experimental data and the evaluated libraries were discussed. This study provides a reliable benchmark for nuclear data evaluation and model optimization, and lays a solid foundation for Rh medical isotope applications and astrophysical research.

\end{abstract}

\end{frontmatter}

\section{Introduction} \label{sec:intro}
\subsection{$^{103}$Rh(n,$\gamma$) cross section}

The slow (s) and rapid (r) neutron capture processes are the primary mechanisms responsible for the synthesis of heavy elements in stars~\cite{meyer1994r}. In the s-process, stable isotopes capture neutrons gradually and subsequently undergo $\beta$ decay. The neutron capture rates of these isotopes critically affect the resulting elemental abundances~\cite{gangopadhyay2019nuclear}. Therefore, accurate measurements of MACS in the keV energy range are essential for reliable nuclear astrophysics models and for understanding the origin of heavy elements~\cite{bao2000neutron}. As the only stable isotope of rhodium, $^{103}$Rh plays a significant role in the s-process nucleosynthesis path, as illustrated in Fig.~\ref{sp-process}.

As a high-yield $^{235}$U fission product with strong neutron absorption, $^{103}$Rh is vital in reactor fuel cycle optimization and core safety design~\cite{Reffo1982FastNC}. In particular, rhodium-based self-powered neutron detectors (SPNDs), extensively deployed in nuclear systems such as China’s HPR1000, offer excellent radiation resistance and strong signal output without external power~\cite{XING201679,WU2024105110}. Then, accurate neutron capture cross section measurements are essential for enhancing detection sensitivity. Notably, Rh-SPNDs developed by CNEA have been clinically validated for thermal neutron monitoring in boron neutron capture therapy (BNCT)~\cite{miller2011rhodium}.

The $^{103}$Rh(n,$\gamma$) reaction has been extensively studied, with most efforts focused on the unresolved resonance region (URR, 4--1000~keV), while data in the resolved resonance region (RRR, 0.3--4~keV) remain limited. EXFOR data and evaluated libraries such as ENDF/B-$\mathrm{VIII}$.1 and TENDL-2023 show substantial discrepancies. Lee et al.~\cite{lee2003neutron} obtained detailed structure around 1.26~eV using time-of-flight(TOF) techniques in 2003, but resolution above 50~eV was insufficient, with peak values differing from evaluations by factors of 5--40. Popov et al.~\cite{popov1962energy} also reported low-resolution data, and no other continuous data below 100~eV are available. Between 100~eV and 4~keV, experimental and evaluated results disagree on peak positions and magnitudes. URR data from Carlson~\cite{osti_4022848}, LeRigoleur~\cite{le1976mesures}, Macklin~\cite{macklin1980100}, Bokhovko~\cite{Bokhovko} are more consistent but still vary by 10--40~\%. Consequently, derived MACS values at $kT$ = 30~keV range from $741 \pm 37$ to $1019 \pm 41$~mb, indicating that current datasets remain insufficient to constrain stellar nucleosynthesis models.

In this work, we measured the neutron capture cross section of $^{103}$Rh in the 0.3~eV--1~MeV range using the Back-n facility based on CSNS, obtain resonance structure parameters through the $R$-matrix program, and calculate the cross section in the URR region using TALYS, and derive the corresponding MACS.

\subsection{$^{103}$Rh($\gamma$,n) cross section}

\begin{figure}[t]
\centering
\includegraphics[width=1.0\linewidth]{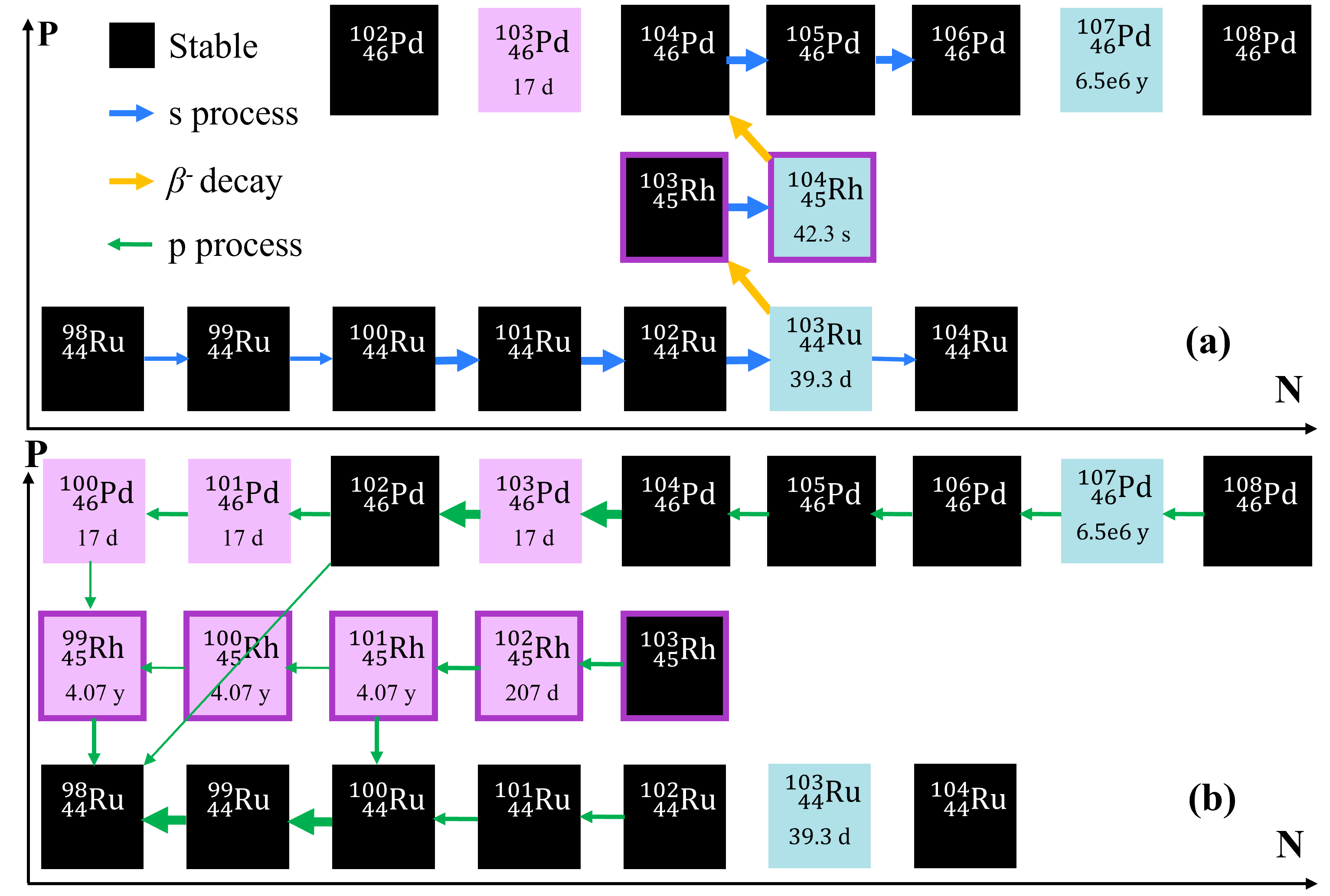}
\caption{\label{sp-process}(a) The main s-process for $^{103}$Rh in massive stars.(b) The p-process for core-collapse supernova around $^{103}$Rh.}
\end{figure}

$^{103}$Rh, a key material in nuclear technology, is crucial for photoneutron reaction studies. In the astrophysical p-process, which includes two mechanisms: photoneutron reaction ($\gamma$, n) and proton capture (p, $\gamma$), the $^{103}$Rh($\gamma$, n) reaction is one of the critical steps, as illustrated in Fig.~\ref{sp-process}(b), significantly influencing the predicted abundances of p-nuclide isotopes (e.g., $^{98}$Ru). The production cross section of $\rm^{102m}$Rh and $\rm^{102g}$Rh can validate nuclear models and optimize production conditions~\cite{shakilur2016measurement,KAWANO2020109}. The $^{103}\mathrm{Rh}(\gamma,\text{n})^{102}\mathrm{Rh}$ and $^{103}\mathrm{Rh}(\gamma,\gamma')^{103\mathrm{m}}\mathrm{Rh}$ reactions share the same initial excited state. An increase in the cross section of the $(\gamma,\text{n})$ channel can significantly suppress the yield of the isomeric state $^{103\mathrm{m}}\mathrm{Rh}$, whereas the $(\gamma,\gamma')$ channel favors energy storage and subsequent formation of $^{103\mathrm{m}}\mathrm{Rh}$. The ratio of the cross section between the two channels directly affects the isomeric yield. Given that $^{103\mathrm{m}}\mathrm{Rh}$ emits Auger electrons via internal conversion, it holds great potential for targeted radiotherapy and SPECT molecular imaging~\cite{PAN2021109534,pang2023progress}. A quantitative comparison of the two cross section provides essential experimental data for optimizing $^{103\mathrm{m}}\mathrm{Rh}$ production conditions and improving its yield.

Experimental data on the $^{103}$Rh($\gamma$,n)$^{102}$Rh reaction show notable inconsistencies. Lepretre et al.~\cite{LEPRETRE197439} observed the $^{103}\mathrm{Rh}$ GDR peak at Saclay in 1974 using the positron annihilation-in-flight technique (16.15~MeV, 191~mb), but the peak shape exhibited clear oscillations, limiting the precision of both the peak energy and width. Varlamov et al.~\cite{varlamov2019evaluation} questioned the reliability of the Saclay $\sigma(\gamma,\mathrm{in})$ data (i=1,2,3) on physical consistency grounds. Goriely et al.~\cite{PhysRevC.102.064309} measured a GDR peak via laser Compton scattering (16.55~MeV, 193~mb) at NewSUBARU, with a much clearer shape and higher credibility. Although their peak magnitudes are similar, the two datasets differ in peak position, shape and fine structure. Below 13~MeV, the NewSUBARU cross section are generally lower than those from Saclay; above 13~MeV, they are generally higher. Their maximum discrepancy of about 10\%. Based on the NewSUBARU data, IFIN-NH employed the EMPIRE model in 2019 to produce the IAEA/PD-2019 evaluation, which agrees extremely well with the original NewSUBARU measurements in 9--17~MeV. In 2024, IFIN-NH updated this evaluation for ENDF/B-$\mathrm{VIII}$.1. In 15--17~MeV, the ENDF/B-$\mathrm{VIII}$.1 values lie slightly below IAEA/PD-2019--which means, below NewSUBARU but above Saclay. TENDL-2023, however, lies significantly higher than all of the above in 10--13.5~MeV (up to $\sim21\%$ above Saclay and $\sim37\%$ above NewSUBARU), and in 13.5--17~MeV it is close to Saclay but lower than the others. Hence, these datasets exhibit marked differences in GDR peak energy, width, and intensity.

Employing the innovative LCS source available at SLEGS, we are able to produce energetic, almost single-color $\gamma$-rays. This will significantly aid investigations involving ($\gamma$,n) reactions. Using this new approach promises to reconcile inconsistencies found in previous results and lead to more precise and dependable measurements of $^{103}$Rh($\gamma$,n) reaction cross section.

\section{Measurement of the $^{103}$Rh(n,$\gamma$) cross section}
\subsection{Experiment}

The China Spallation Neutron Source (CSNS) employs a 1.6~GeV pulsed proton beam at 25~Hz to bombard a tungsten target. The back streaming white neutron facility (Back-n) at CSNS, operational since 2018, provides a neutron energy spectrum of 0.1~eV--400~MeV with a flux of 10$^{7}$~n/cm$^{2}$/s and TOF capabilities~\cite{PhysRevResearch.6.013225,tang2021back}. In this work, Back-n operated in double-bunch mode with a 41~ns FWHM of proton pulse width and 410 ns interval, conducted at ES\#2, 76 meters from the target, thereby enhancing time resolution and reducing $\gamma$-ray interference. A cadmium filter was used to absorb low-energy neutrons, while two collimators shaped the beam into a 40 mm Gaussian profile.

The $\rm{C_{6}D_{6}}$ detectors, with low neutron sensitivity, use Pulse Height Weighting Technique (PHWT) to eliminate dependence on cascading $\gamma$-ray energy. The EJ315 scintillator and photomultiplier tube enable a 10~ns response, with signals acquired by a PXIe-based system offering 12-bit resolution and 1~GS/s sampling~\cite{ren2019c}.

Four target materials were used in this work: $^{103}$Rh, $\rm^{nat}C$, $\rm^{nat}Pb$, and $\rm^{197}Au$. $^{103}$Rh was the sample under study(diameter 30mm, thickness 1mm, density 12.30 $\rm{g/cm^3}$, purity 99.95\%, impurity content shown in Table.~\ref{impurity}), $\rm^{nat}C$ and $\rm^{nat}Pb$ subtracted background from scattered neutrons and in-beam $\gamma$-rays, and $\rm^{197}Au$ calibrated the neutron flight path and detection system. Following irradiation, sufficient statistics were acquired for reliable analysis. Additionally, a Ta-Co filter corrected in-beam $\gamma$-ray background using the black resonance method~\cite{an2023measurement}. 
\begin{table}[htbp]
\centering
\footnotesize
\caption{\label{impurity}Impurity content of Rh target in ppm (1 ppm = 0.0001\%)}
\begin{tabular}{lccccccccccc}
\hline
Au & Ag & Pt & Pd & Ru & Ir & Fe & Cu & Ni & Al & Pb \\
\hline
13 & 4 & 15 & 13 & 16 & 10 & 12 & 8 & 13 & 10 & 3 \\
\hline
\end{tabular}
\end{table}

\vspace{-0.5cm}

\begin{figure}[h!]
\centering
\includegraphics[width=0.99\linewidth] {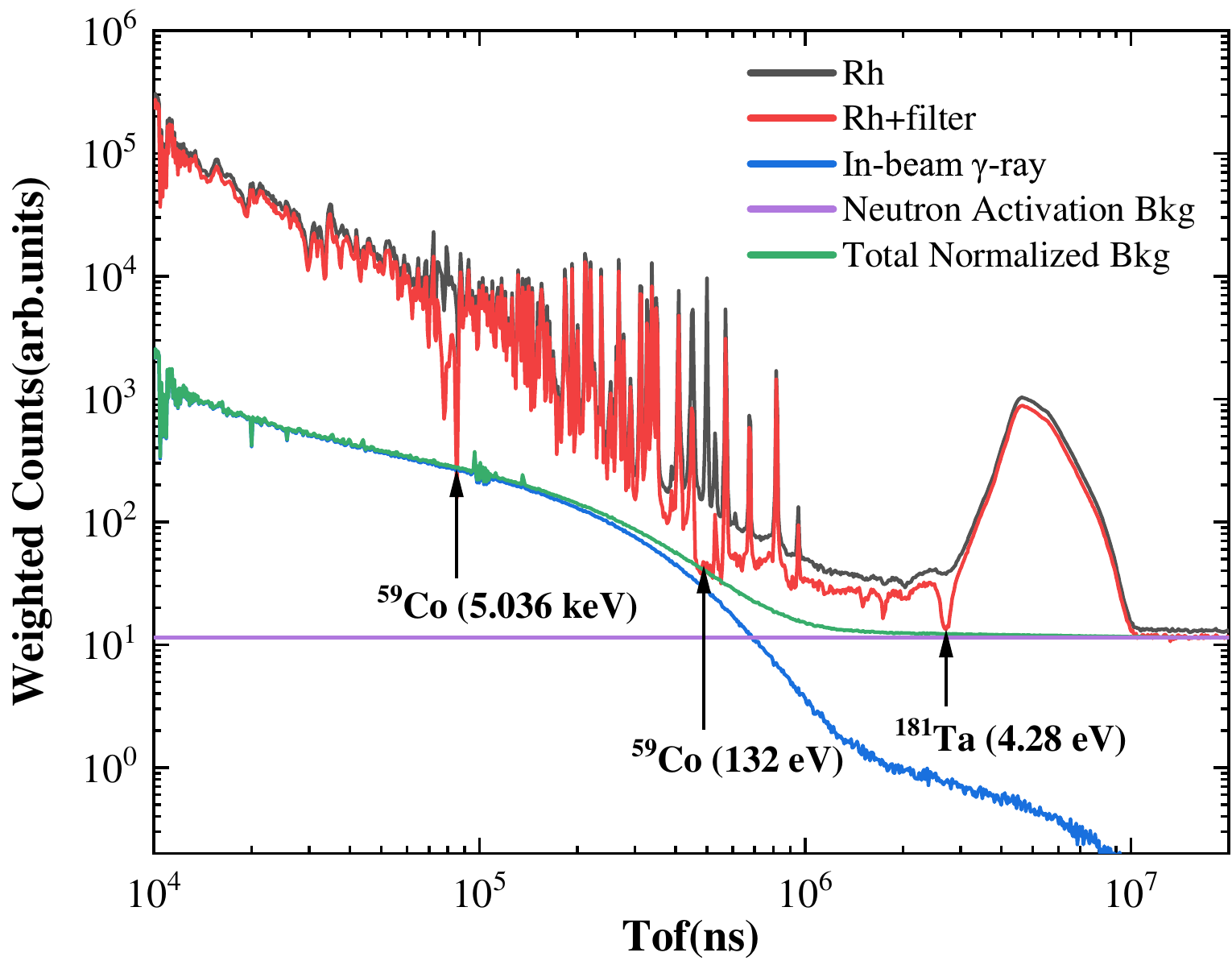}
\caption{\label{background}Schematic of the background compositions of $^{103}$Rh.}
\end{figure}

\subsection{Data Analysis}

The experimental background primarily includes the sample-related background \( B_{\text{sample}}(t_n) \) and the sample-independent background \( B_{\text{empty}}(t_n) \). The sample-related background \( B_{\text{sample}}(t_n) \) consists of the induced background from scattered neutrons \( B_{\text{sn}}(t_n) \), the background from scattered in-beam $\gamma$-rays \( B_{\text{s}\gamma}(t_n) \), and the background from sample activation. The sample-independent background \( B_{\text{empty}}(t_n) \) is determined using an empty target. The \( B_{\text{sn}}(t_n) \) is estimated using counts from a carbon target, while \( B_{\text{s}\gamma}(t_n) \) is obtained through experiments with a lead target and a sample combined with an absorber. The latter is derived by normalizing to the absorption dip. The raw count spectra for each target and the in-beam $\gamma$-ray background are shown in Fig.~\ref{background}.

\begin{figure*}[htbp]
\centering
\includegraphics[width=0.92\linewidth] {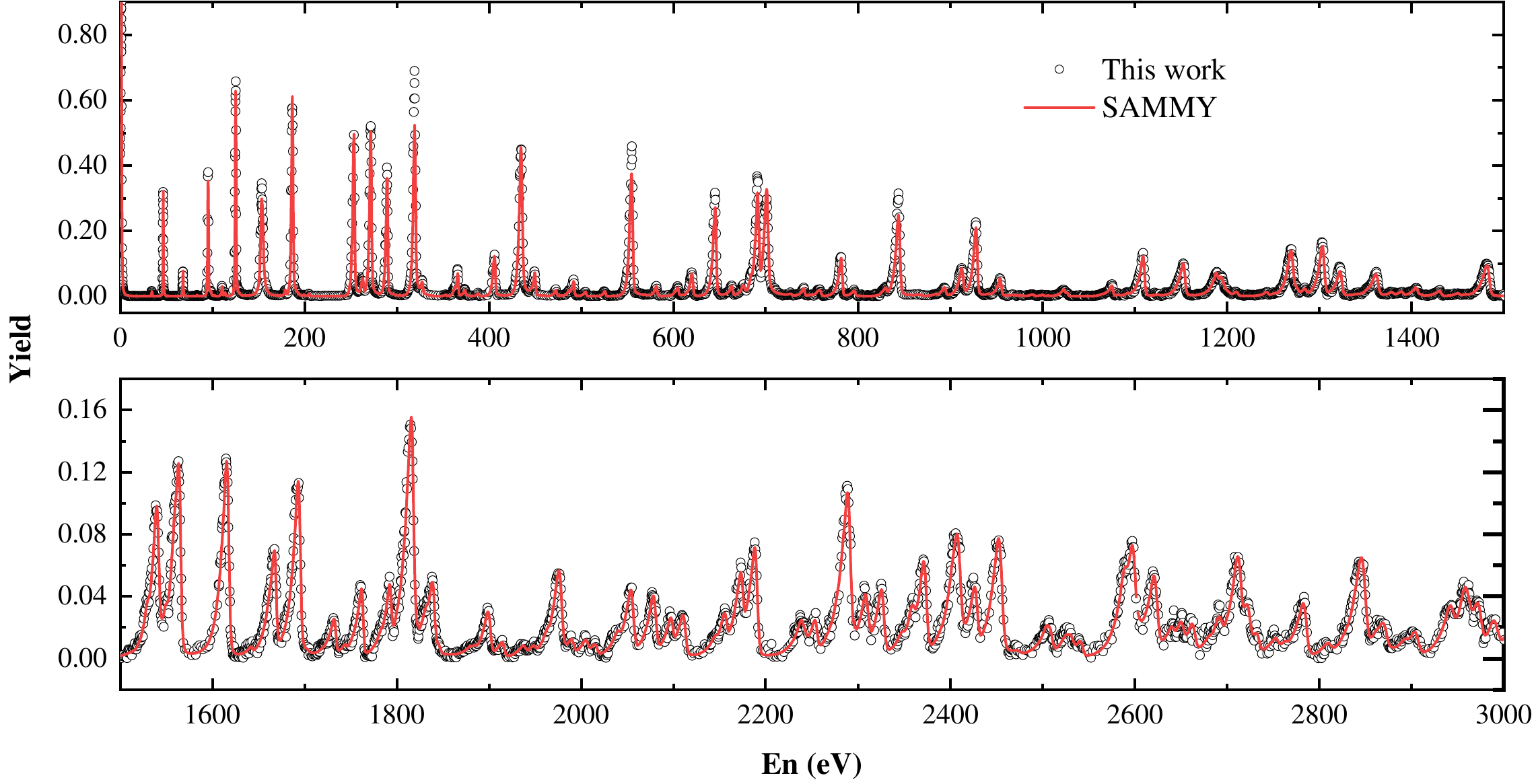}
\caption{\label{Yield}Neutron capture yield of this work and SAMMY fit in the range of 0-3000~eV}
\end{figure*}

\begin{figure*}[htbp]
\centering
\includegraphics[width=0.93\linewidth]{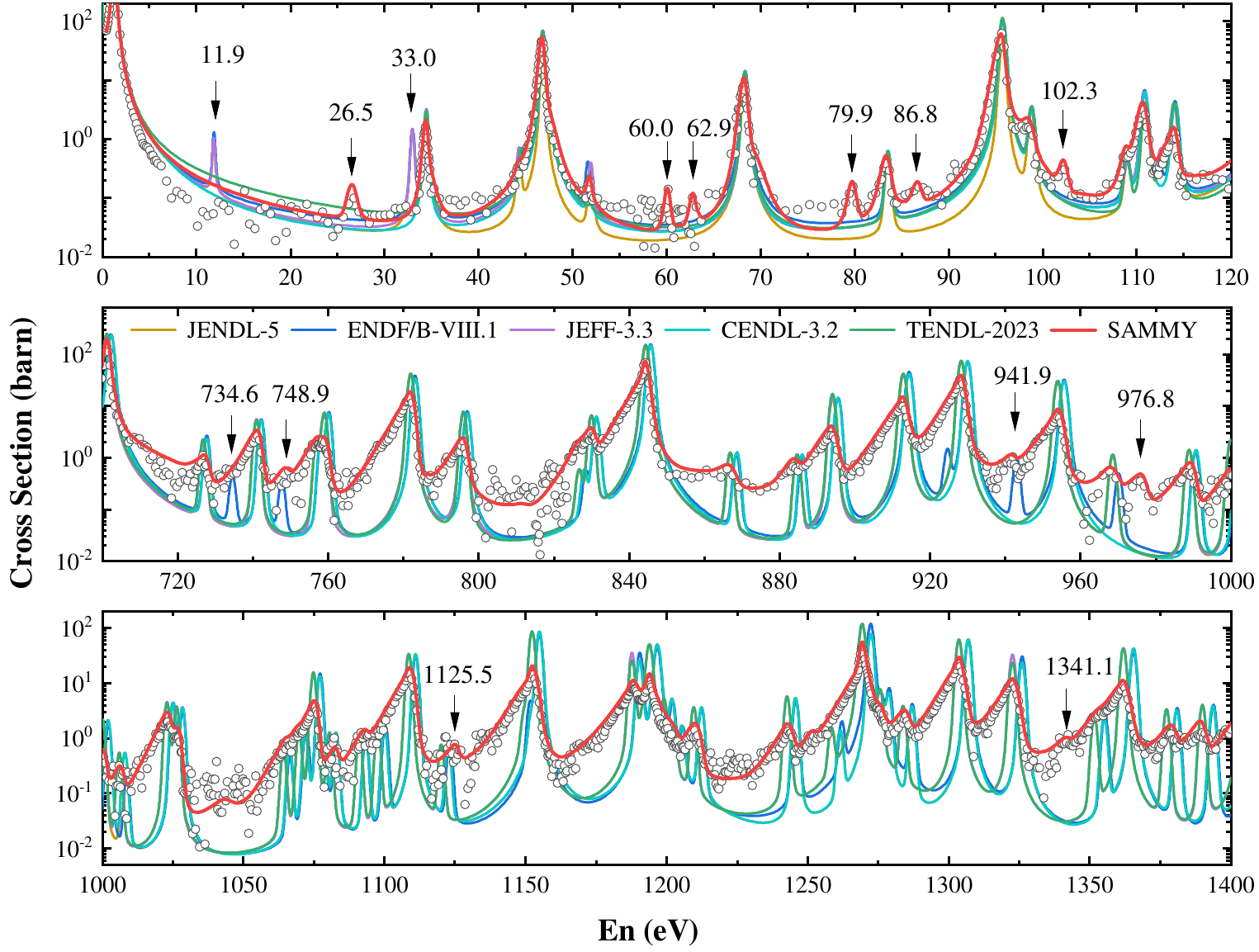}
\caption{\label{RRR}The experimental cross section were compared with evaluation data in the range of 0-1400~eV.}
\end{figure*}

The $\rm{C_{6}D_{6}}$ detector exhibits excellent neutron insensitivity and lower detection efficiency, allowing the assumption that its detection efficiency is approximately equal to the sum of the cascade $\gamma$-ray detection efficiency in capture reactions. PHWT was thus developed ~\cite{borella2007use,PhysRevC.108.035802}. 

After applying PHWT, the capture yield ($Y_w$) can be determined. The relationship between the neutron capture cross section  $\sigma_c$ and the reaction yield is as follows:
\begin{equation}
\sigma_c=Y_w(E)\frac{\sigma_t(E)}{1-e^{-N\sigma_t(E)tf_c}},
\label{eq:9}
\end{equation}
where $\sigma_t$ is the neutron total cross section, $N$ is the atom density, and $t$ is the target thickness. Due to the multiple scattering effect in a thick target, a correction factor $f_c$ is introduced in calculation and is ascertained through GEANT4 simulations.

The uncertainties primarily arise from experimental conditions and data analysis. In terms of experimental conditions, uncertainties include those related to sample parameters, neutron spectrum, and proton beam power. The uncertainty in sample parameters is reflected in the atomic surface density of the sample. According to Chen et al.~\cite{Chen2019}, who analyzed the neutron spectrum of Back-n using a Li-Si detector, the uncertainty in the spectrum is less than 8.0\% below 0.15~MeV and less than 4.5\% above 0.15~MeV. Additionally, the uncertainty in the proton beam power of the CSNS was less than 1.5\%. In terms of data analysis, uncertainties mainly stem from the PHWT and the double bunch spectrum unfolding method. According to Tain et al.~\cite{Tain2002}, the systematic error introduced by the PHWT is approximately 2.0\%--3.0\%. Then, the total uncertainty in the results is less than 10.0\%. 

\subsection{Result and discussion} \label{sec:ese}

For the RRR, the neutron capture yield measured in this study was fitted using the $R$-matrix code SAMMY~\cite{larson2008updated}, with input parameters primarily taken from the TENDL-2023 evaluation. The fitting accounted for experimental effects such as Doppler and resolution broadening, self-shielding, and multiple scattering~\cite{yang2023measurement,PhysRevC.100.045804}. Fig.~\ref{Yield} shows a portion of the SAMMY-fitted neutron capture yield. The fitted curve agrees well with the experimental data, indicating good fitting quality. Overall, the cross section data align well with major evaluations, with several new structures and discrepancies identified, as illustrated in Fig.~\ref{RRR}.

\begin{figure}[htbp]
\centering
\includegraphics[width=0.96\linewidth]{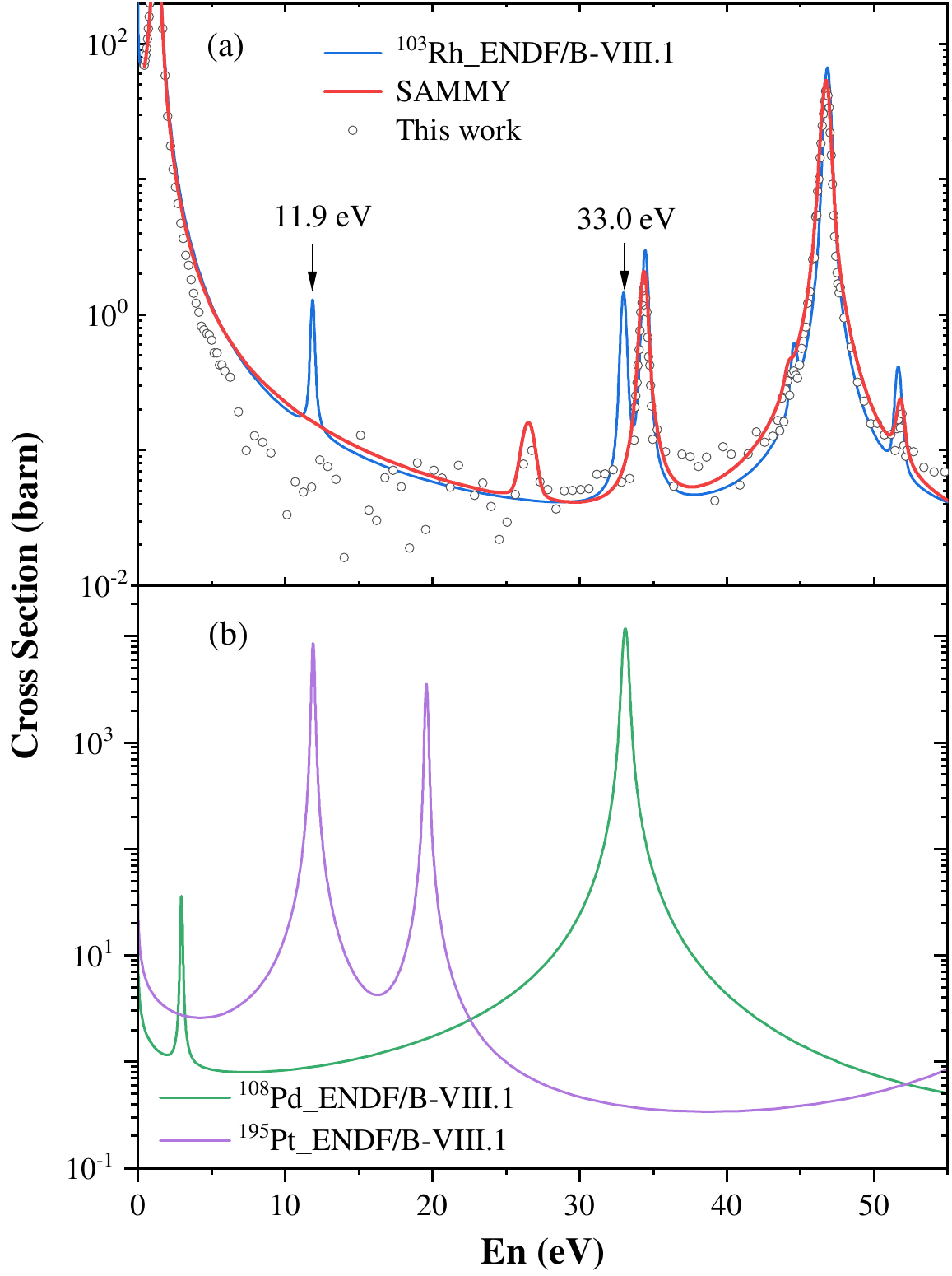}
\caption{\label{pd108pt195}The comparison of $^{103}\text{Rh}$, $^{108}\text{Pd}$ and $^{195}\text{Pt}$ cross section.}
\end{figure}

In this work, no evidence for the 11.9~eV and 33.0~eV resonance structure reported in ENDF/B-$\mathrm{VIII}$.1, JEFF-3.3, and JENDL-5 was observed, with these two resonance parameters originating from measurements by Brusegan et al.~\cite{brusegan2005neutron}. Rhodium is a platinum-group metal whose chemical properties and crystal structure are very similar to those of its peers, making isomorphous substitution easy to occur in nature and leading to mutual replacement within mineral lattices. As a result, rhodium crystals typically contain other platinum-group metals, such as Ru, Pt and Pd etc. By comparing the capture cross section of these platinum-group metals, we found that the 11.9 eV resonance most likely arises from $^{195}$Pt, whereas the 33.0 eV resonance most likely arises from $^{108}$Pd~\cite{brusegan2005neutron}, as shown in Fig~\ref{pd108pt195}. In the present work, Rh has a purity of 99.95\%, with Pt and Pd impurities of only 0.0015\% and 0.0013\%, respectively, thereby eliminating the resonance structures associated with these impurities. 

New resonance structures can be clearly observed at 26.5, 79.9, 86.8, 102.3, 941.9, and 976.8~eV, which are not provided by any evaluated libraries or previous measurements. The structure at 941.9~eV further validates the data from ENDF/B-$\mathrm{VIII}$.1. At 60.2 and 62.9~eV, the original count spectrum shows distinct structures; however, the cross section values here are small, and although the data processing errors are relatively large compared to these values, the statistical results are poor, making it impossible to present clear cross section structures. These two points may represent new resonance structures but require more precise experimental analysis. Additionally, ENDF/B-$\mathrm{VIII}$.1 indicates resonance structures at 734.6 and 748.9~eV. Although the experimental values here do not show obvious peaks, small protrusions are still present, partially validating these two structures. Around 1122.0~eV, there is a slight difference in the peak energy positions between ENDF/B-$\mathrm{VIII}$.1 and TENDL-2023 (1120.1~eV and 1123.9~eV, respectively), and our experimental results show the peak at 1125.5~eV. At 1341.1~eV, there is a clear structure, but due to the influence of the peak on the right, a distinct peak cannot be fitted. However, it is clear that a resonance structure exists here.

\begin{figure}[htbp]
    \centering
    \includegraphics[width= \linewidth]{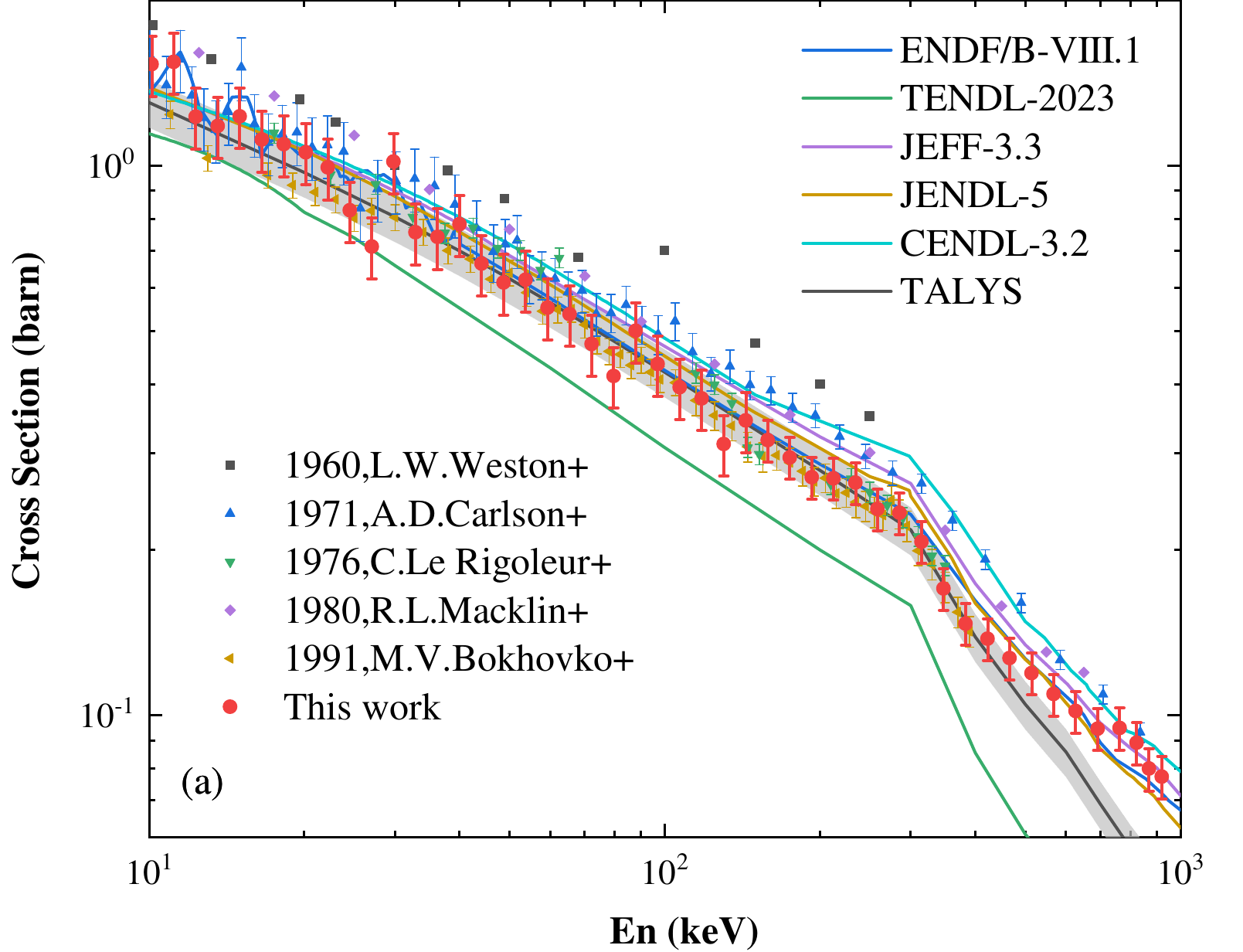}
    \includegraphics[width= \linewidth]{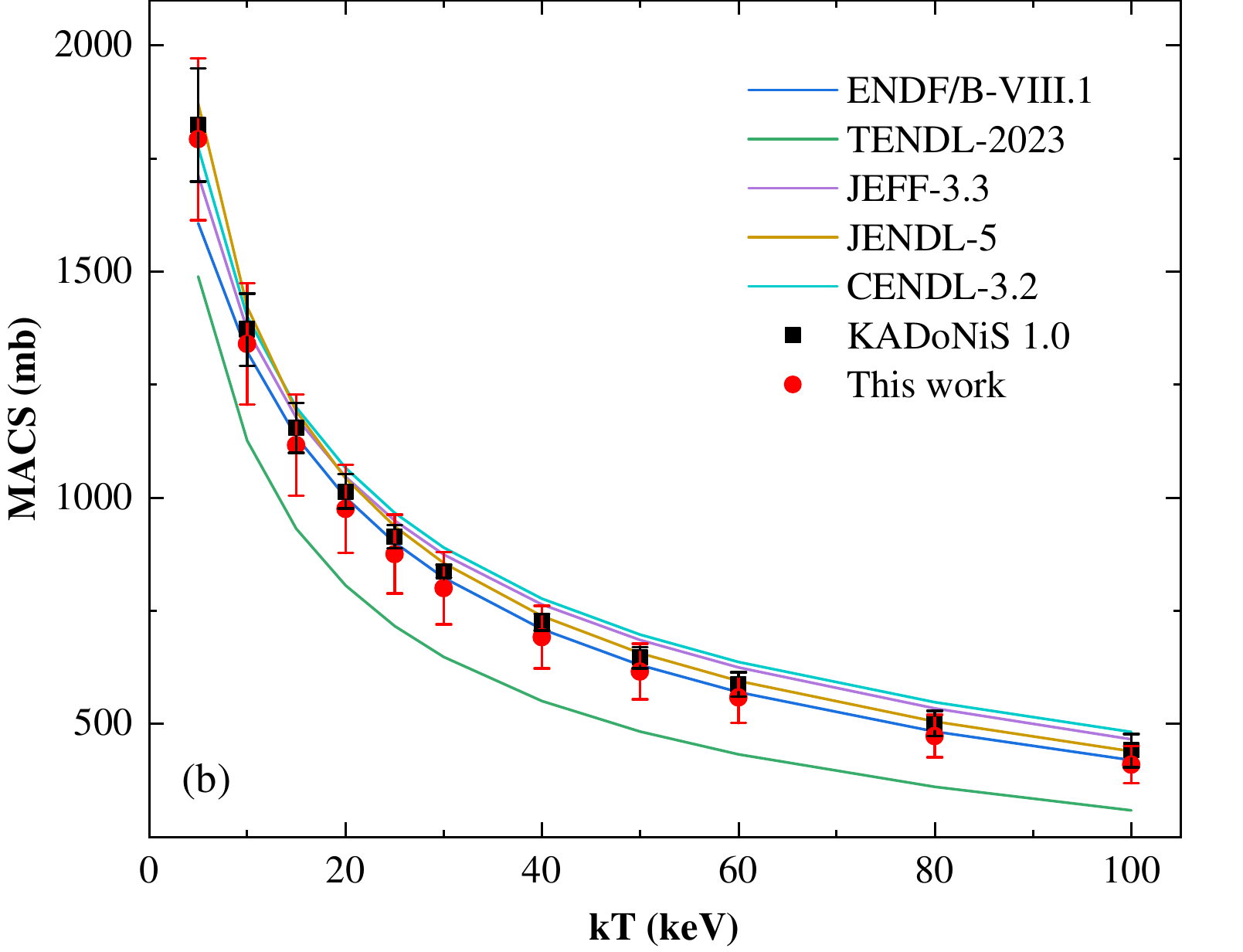}
    \includegraphics[width= \linewidth]{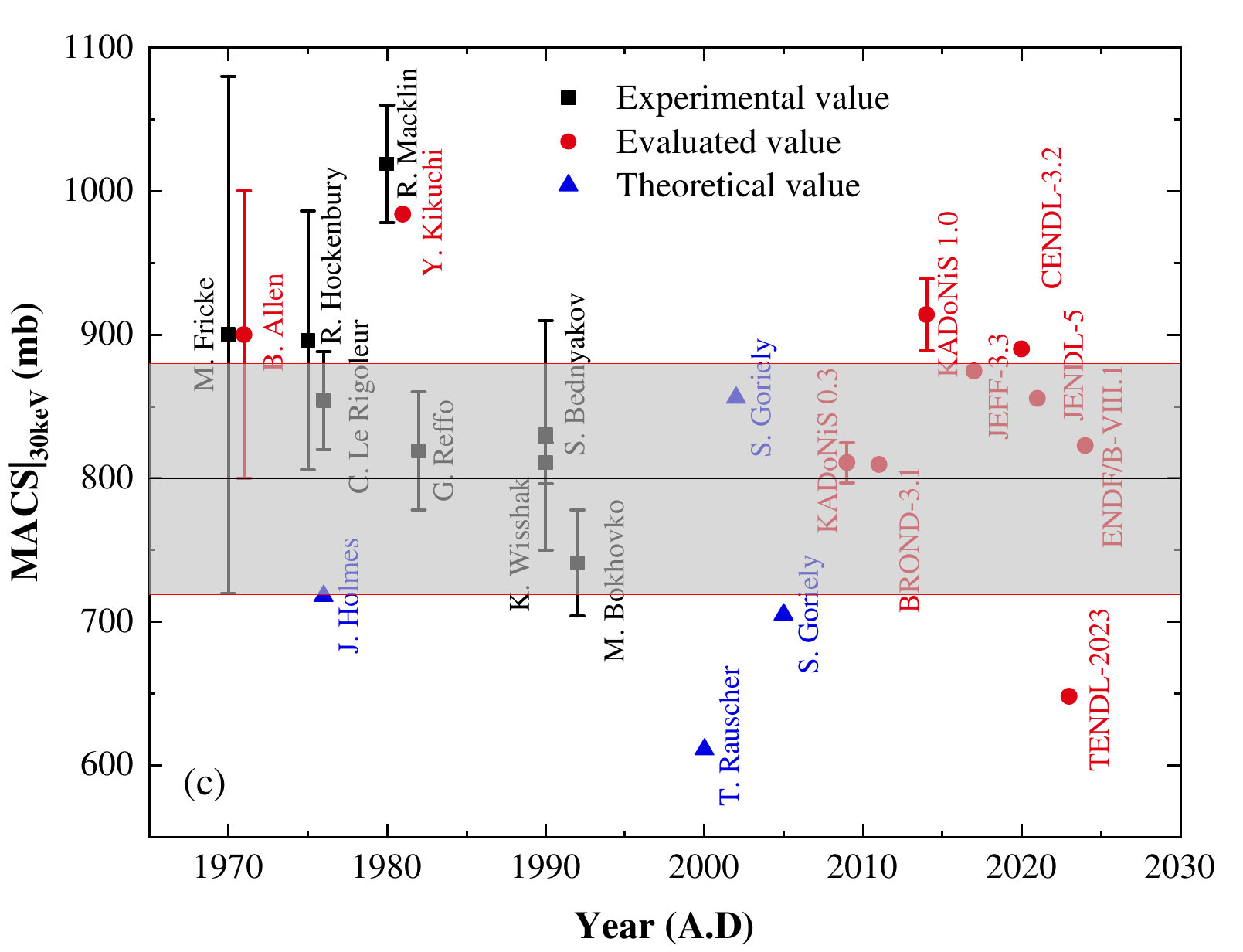}
    \caption{\label{URR-MACS}(a) The neutron capture cross section of \(^{103}\)Rh in the URR from 10 to 1000~keV. The grey band represents the 10\% uncertainty of TALYS. (b) Comparison of the experimental MACS from $kT$ = 5--100~keV with evaluated data and KADoNiS 1.0, and (c) MACS at thermal energy $kT$ = 30~keV.}
\end{figure}

The average neutron capture cross section in the URR between 10–900~keV obtained in this work is shown in Fig.~\ref{URR-MACS}(a) (tabulated in Table.~\ref{table_URR}), compared with major evaluated libraries and previous experimental data. Overall, the results of this work align closely with the ENDF/B-$\mathrm{VIII}$.1 data, with oscillations in the 10--40~keV range partially reflected. The JEFF-3.3 and JENDL-5 data are higher, while the TENDL-2023 data are lower. The results of this work in the 10--40~keV range are consistent with those from Carlson et al. and LeRigoleur et al., while in the 40--1000~keV range, they align closely with Bokhovko et al. However, the data from Weston, Carlson and Macklin et al. are higher. The TALYS program describes the average cross section of isotopes in the URR~\cite{Koning2023,li2022measurement}. The theoretical cross section will be further used to calculate the MACS of the isotope.

The obtained MACS for $^{103}$Rh is shown in Fig.~\ref{URR-MACS}(b)(tabulated in Table.~\ref{table_MACS30}), compared with multiple evaluated libraries and KADoNiS 1.0 data. Overall, the MACS from this study follows a trend consistent with the data from JENDL-5 and KADoNiS 1.0 (which nearly overlap), though slightly lower. Apart from the point at 5~keV, it aligns closely with the evaluated data from ENDF/B-$\mathrm{VIII}$.1. In contrast, the evaluated values provided by TENDL-2023 appear significantly underestimated, while those from JEFF-3.3 show slight deviations but remain relatively close.

\begin{table}[htbp]
\centering
\footnotesize
\begingroup
\caption{\label{table_URR}The (n,$\gamma$) cross section and the uncertainties of $^{103}\text{Rh}$ in the URR.}
\begin{tabular}{ccc|ccc}
\hline
\makecell{$E_{\text{low}}$\\(keV)} & \makecell{$E_{\text{high}}$\\(keV)} & \makecell{$\sigma_c$\\(barn)} & \makecell{$E_{\text{low}}$\\(keV)} & \makecell{$E_{\text{high}}$\\(keV)} & \makecell{$\sigma_c$\\(barn)} \\
\hline
10.6 & 11.7 & 1.544 $\pm$ 0.195 & 101.8 & 112.3 & 0.395 $\pm$ 0.050 \\
11.7 & 12.9 & 1.227 $\pm$ 0.156 & 112.3 & 123.9 & 0.377 $\pm$ 0.048 \\
12.9 & 14.3 & 1.180 $\pm$ 0.150 & 123.9 & 136.7 & 0.312 $\pm$ 0.039 \\
14.3 & 15.7 & 1.229 $\pm$ 0.156 & 136.7 & 150.8 & 0.344 $\pm$ 0.043 \\
15.7 & 17.4 & 1.114 $\pm$ 0.141 & 150.8 & 166.4 & 0.316 $\pm$ 0.028 \\
17.4 & 19.2 & 1.093 $\pm$ 0.139 & 166.4 & 183.5 & 0.294 $\pm$ 0.026 \\
19.2 & 21.1 & 1.057 $\pm$ 0.134 & 183.5 & 202.5 & 0.271 $\pm$ 0.024 \\
21.1 & 23.3 & 0.992 $\pm$ 0.125 & 202.5 & 223.4 & 0.269 $\pm$ 0.023 \\
23.3 & 25.7 & 0.829 $\pm$ 0.105 & 223.4 & 246.4 & 0.265 $\pm$ 0.023 \\
25.7 & 28.4 & 0.712 $\pm$ 0.090 & 246.4 & 271.9 & 0.237 $\pm$ 0.021 \\
28.4 & 31.3 & 1.016 $\pm$ 0.129 & 271.9 & 299.9 & 0.233 $\pm$ 0.020 \\
31.3 & 34.6 & 0.756 $\pm$ 0.096 & 299.9 & 330.9 & 0.207 $\pm$ 0.018 \\
34.6 & 38.1 & 0.741 $\pm$ 0.094 & 330.9 & 365.0 & 0.170 $\pm$ 0.015 \\
38.1 & 42.1 & 0.782 $\pm$ 0.099 & 365.0 & 402.7 & 0.147 $\pm$ 0.013 \\
42.1 & 46.4 & 0.663 $\pm$ 0.084 & 402.7 & 444.3 & 0.138 $\pm$ 0.012 \\
46.4 & 51.2 & 0.612 $\pm$ 0.077 & 444.3 & 490.1 & 0.127 $\pm$ 0.011 \\
51.2 & 56.5 & 0.620 $\pm$ 0.078 & 490.1 & 540.7 & 0.119 $\pm$ 0.010 \\
56.5 & 62.3 & 0.552 $\pm$ 0.070 & 540.7 & 596.5 & 0.109 $\pm$ 0.010 \\
62.3 & 68.7 & 0.536 $\pm$ 0.068 & 596.5 & 658.1 & 0.102 $\pm$ 0.009 \\
68.7 & 75.8 & 0.474 $\pm$ 0.060 & 658.1 & 726.0 & 0.094 $\pm$ 0.008 \\
75.8 & 83.6 & 0.414 $\pm$ 0.052 & 726.0 & 792.2 & 0.095 $\pm$ 0.008 \\
83.6 & 92.3 & 0.500 $\pm$ 0.063 & 792.2 & 846.2 & 0.089 $\pm$ 0.008 \\
92.3 & 101.8 & 0.435 $\pm$ 0.055 & 846.2 & 894.8 & 0.080 $\pm$ 0.007 \\
\hline
\end{tabular}
\endgroup
\end{table}

Fig.~\ref{URR-MACS}(c) presents a comparison of this study's results with the recommended values at $kT$ = 30~keV. Earlier experimental and evaluated values were relatively high, while the results of this study align with the later evaluations, theoretical values, and experimental data. Notably, they exhibit excellent agreement with KADoNiS 0.3 and ENDF/B-$\mathrm{VIII}$.1, while the values from KADoNiS 1.0 are comparatively higher.

\setlength{\tabcolsep}{4pt}
\begin{table}[htbp]
\footnotesize
\centering
\caption{\label{table_MACS30}MACS calculated for this work in the energy range $kT = 5-100$~keV, compared with KADoNiS 0.3 and KADoNiS 1.0.}
\begin{tabular}{cccc}
\toprule
\begin{tabular}[c]{@{}c@{}}$kT$ (keV)\end{tabular} &
\begin{tabular}[c]{@{}c@{}}This work (mb)\end{tabular} &
\begin{tabular}[c]{@{}c@{}}KADoNiS 0.3 (mb)\end{tabular} &
\begin{tabular}[c]{@{}c@{}}KADoNiS 1.0 (mb)\end{tabular} \\
\midrule
5   & $1792 \pm 179$  & 1890             & $1824 \pm 125$ \\
10  & $1339 \pm 134$  & 1392             & $1372 \pm 80$  \\
15  & $1116 \pm 112$  & 1150             & $1154 \pm 55$  \\
20  & $974 \pm 97$    & 998              & $1014 \pm 38$  \\
25  & $875 \pm 88$    & 892              & $914 \pm 25$   \\
30  & $799 \pm 80$    & $811 \pm 14$     & $837 \pm 14$   \\
40  & $691 \pm 69$    & 695              & $725 \pm 19$   \\
50  & $615 \pm 62$    & 616              & $646 \pm 23$   \\
60  & $557 \pm 56$    & 556              & $587 \pm 27$   \\
80  & $472 \pm 47$    & 472              & $501 \pm 28$   \\
100 & $409 \pm 41$    & 413              & $441 \pm 37$   \\
\bottomrule
\end{tabular}
\end{table}
\section{Measurement of the $^{103}$Rh($\gamma$,n) cross section}
\subsection{Experiment}

The Shanghai Laser Electron Gamma Source (SLEGS) is a $\gamma$-ray facility based on inverse Compton scattering of a 10.64~$\mu$m CO$_2$ laser (100~W output power) with a 3.5~GeV electron beam from the SSRF storage ring~\cite{wang2022commissioning}. It operates in back scattering (180°) and slant scattering (20°--160°) modes, generating near-monoenergetic $\gamma$-rays spanning 0.66--21.7~MeV~\cite{7370951}. $\gamma$-rays are collimated by a dual collimator system consisting of a coarse collimator and a fine collimator, ensuring the energy resolution~\cite{XU2025170249}. SLEGS is well-suited for photonuclear cross section measurements, fusion diagnostics, dose assessments, and activation studies, particularly in the GDR region~\cite{li2025new}.

In the $^{103}$Rh($\gamma$,n)$^{102}$Rh experiment, SLEGS provided quasi-monoenergetic $\gamma$-rays covering the 9.32--16.76~MeV range, avoiding double neutron emission. The beam passed through a coarse (C5, 5~mm) and fine (T2, 2~mm) collimator before irradiating a $^{103}$Rh target(diameter 10 mm, thickness 2.38 mm, density 11.73 $\rm{g/cm^3}$, purity 99.95\%, same impurity content shown in Table.~\ref{impurity}) at the center of a $^{3}$He flat-efficiency detector (FED) array. Emitted neutrons were recorded by the FED, while transmitted $\gamma$-rays were attenuated by a copper absorber and measured by a BGO detector to reconstruct the incident energy spectrum.

\begin{figure}[t!]
\centering
\includegraphics[width=0.95\linewidth]{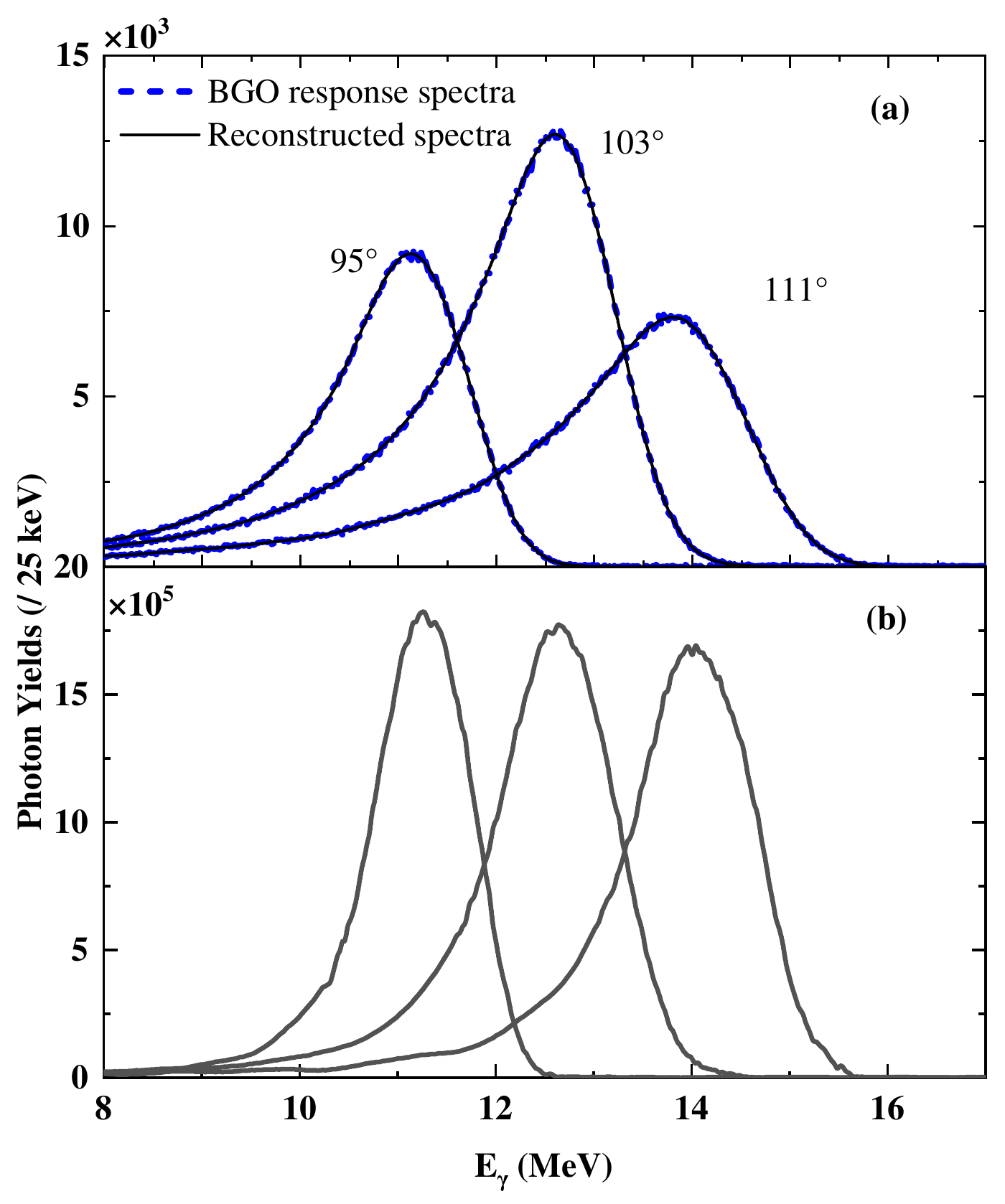}
\caption{\label{Gamma_spectrum} (a) Typical spectra measured by the BGO detector (blue dashed line) and reconstructed from inciden $\gamma$-ray spectra (black solid line). (b) Corresponding $\gamma$ spectra incident on the target.}
\end{figure}

A BGO detector with dimensions 76~mm~$\times$~200~mm was employed to measure the $\gamma$-ray spectra after copper attenuation post-irradiation of the $^{103}$Rh target. Fig.~\ref{Gamma_spectrum} provides examples of $\gamma$-ray spectra detected at slant scattering angles ($\theta_L$) of 95°, 103°, and 111°. To deduce the $\gamma$-ray spectra before the target interaction, a direct deconvolution technique was utilized, combined with the BGO detector's known response function, obtained via GEANT4 simulations~\cite{LIU2024169314}. The reconstructed spectra demonstrated consistency with the detector’s measured spectra, affirming the accuracy of the pre-target $\gamma$-ray energy distribution.

In the experiment, neutrons from the ($\gamma$,n) reaction were detected using a calibrated array of 26 $^{3}$He proportional counters embedded in a polyethylene moderator (450~mm~$\times$~450~mm~$\times$~550~mm)~\cite{HAO20204pi}. The counters are arranged in three concentric rings at 6.5, 11.0, and 17.5~cm from the beam axis, with 1-inch tubes in the inner ring and 2-inch tubes in the middle and outer rings~\cite{jiao2025measurements}. All tubes are 500~mm long, filled with $^{3}$He at 2~atm. Background suppression was achieved using cadmium-loaded polyethylene shielding and a 50~$\mu$s laser duty cycle in a 1000~$\mu$s period. Neutron counts \( N_n \) were extracted via time-normalized background subtraction~\cite{hao2025photoneutron}. The average neutron energy was determined using the ring-ratio technique proposed by Berman and Fultz~\cite{PhysRev.162.1098,RevModPhys.47.713}, calibrated via GEANT4-based efficiency curves. The detector showed high and nearly flat efficiency for neutron evaporation spectra, 1~keV--4~MeV, the typical range for ($\gamma$,n) reactions, which made it suitable for detecting neutrons. The measured efficiency ratio curve agrees well with simulations~\cite{HAO2025}.

\subsection{Data Analysis}

The measurement of photonuclear cross section is based on the interaction between incident $\gamma$-rays and the target material, and the experimental quasi-monoenergetic cross section is expressed as~\cite{PhysRevC.19.1684,PhysRevC.90.064616}:
\begin{equation}{\label{sig_exp}}
    \sigma_{\exp}^{E_{\max}} = \int_{S_n}^{E_{\max}} n_{\gamma}(E_{\gamma}) \sigma(E_{\gamma}) \, \mathrm{d}E_{\gamma} = \frac{N_n}{N_{\gamma} N_t \xi \epsilon_n g},
\end{equation}
where $S_n$ is the single neutron separation energy, $E_{\gamma}$ represents the discrete $\gamma$-ray energy, $n_{\gamma}(E_{\gamma})$ is the normalized $\gamma$-ray energy distribution, $N_n$ is the detected neutron count, $N_{\gamma}$ denotes the incident $\gamma$-ray count, $N_t$ is the target nucleus number per unit area, $\xi$ is the correction factor for thick targets, $\epsilon_n$ is the neutron detection efficiency, and $g$ is the proportion of gamma flux above the neutron threshold.

Due to the broadening of $\gamma$-ray energy distributions, the directly measured $\sigma_{\exp}$ is often a convolution of the incident $\gamma$-ray energy~\cite{GUTTORMSEN1996371}. Unfolding the monoenergetic cross section $\sigma(E_{\gamma})$ requires deconvolution from the following relation~\cite{PhysRevC.109.014617,PhysRevC.98.054619}:
\begin{equation}
    \sigma_f = \mathbf{D} \sigma.
\end{equation}
Here, the matrix $\mathbf{D}$ represents the $\gamma$-ray energy spectrum distribution. The iterative method is employed to solve it by starting with an initial guess $\sigma^0$. In each iteration, the folded cross section is calculated as:
\begin{equation}
    \sigma_f^i = \mathbf{D} \sigma^i.
\end{equation}
The trial function is updated using the experimental measurement $\sigma_{\exp}$ as follows:
\begin{equation}
    \sigma^{i+1} = \sigma^i + (\sigma_{\exp} - \sigma_f^i).
\end{equation}
The process is repeated until the following convergence condition is satisfied:
\begin{equation}
    \sigma_f^{i+1} \approx \sigma_{\exp}.
\end{equation}

\subsection{Result and discussion}
\begin{figure}[t]
\centering
\includegraphics[width=0.95\linewidth]{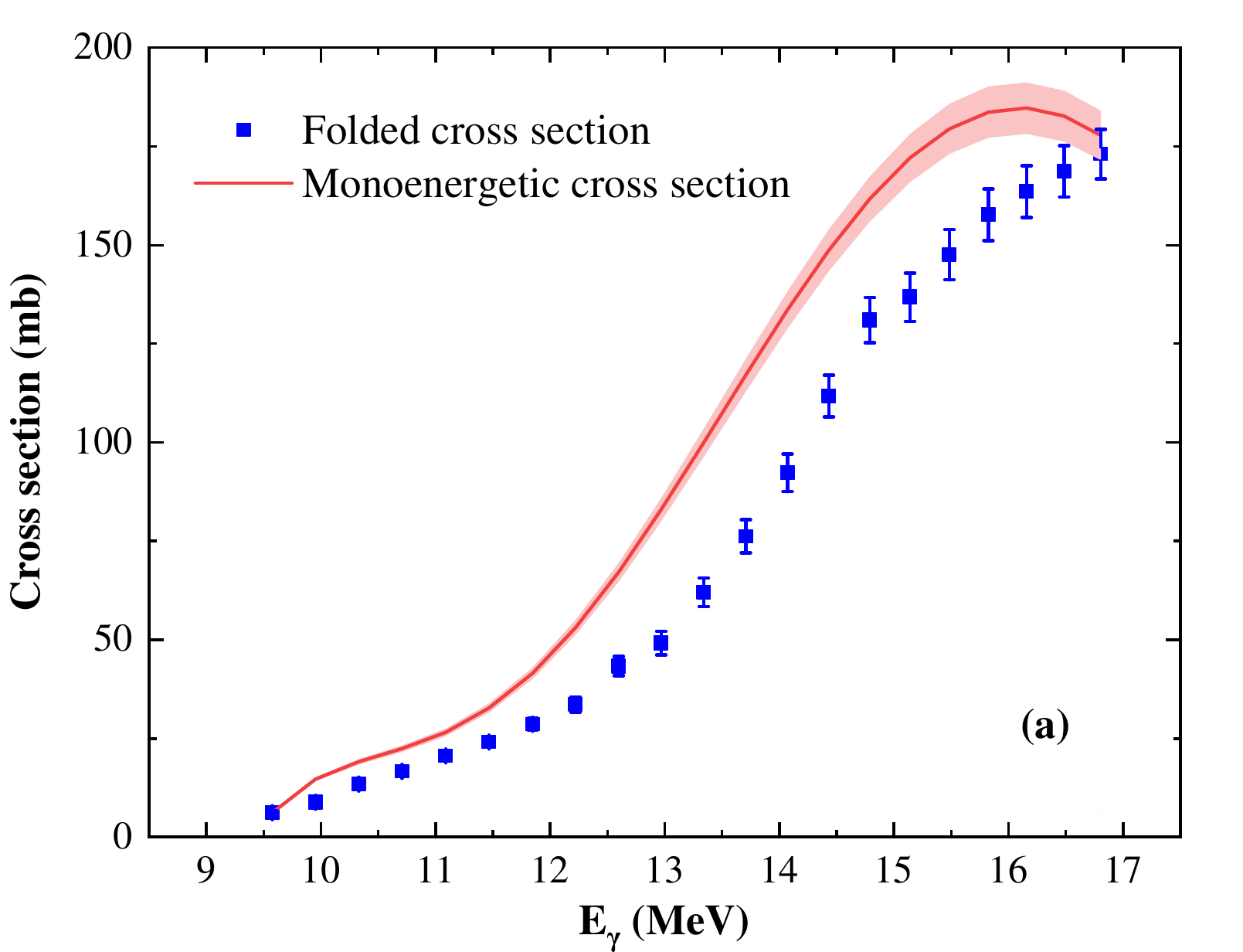}
\includegraphics[width=0.95\linewidth]{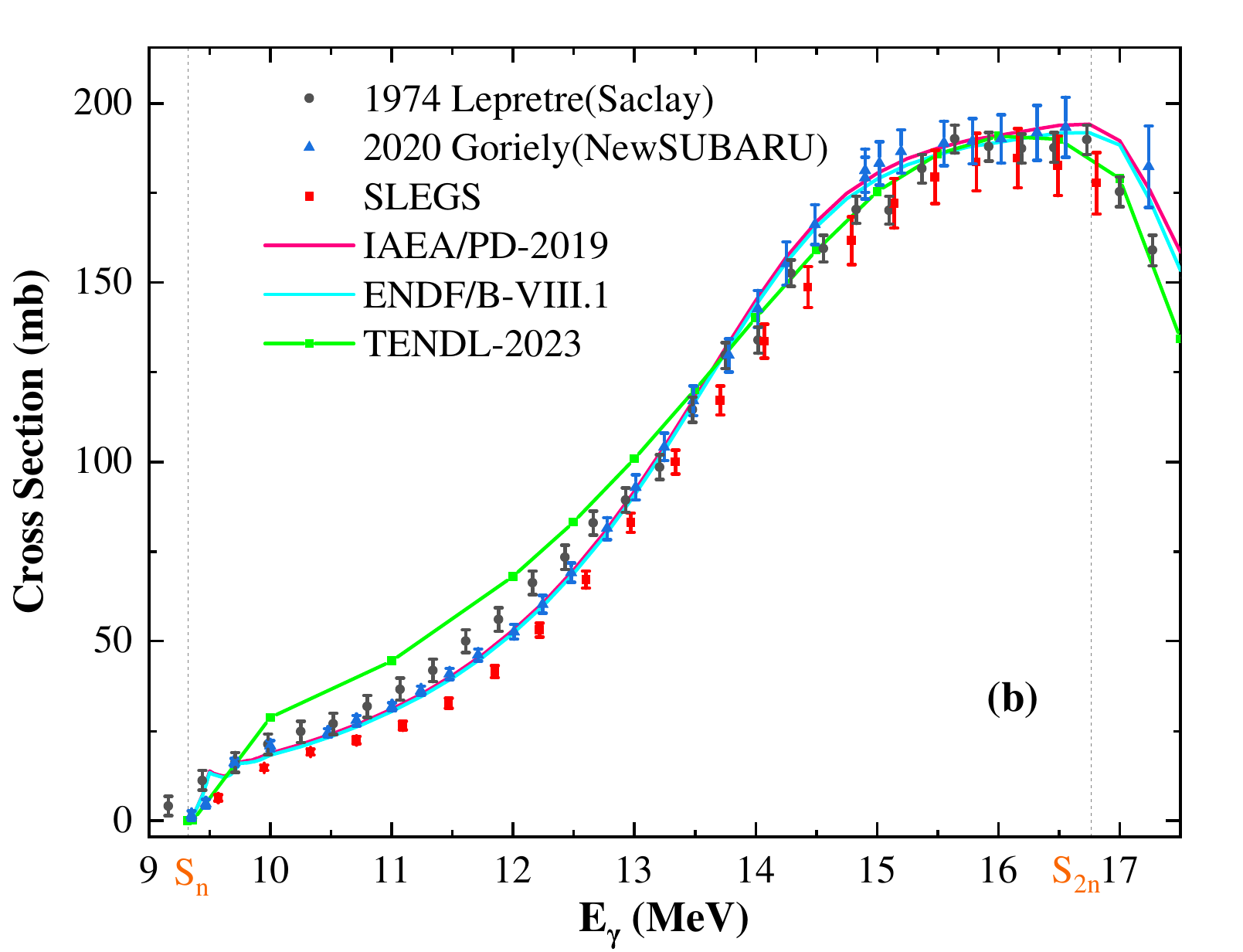}
\caption{\label{Folded} (a) $^{103}$Rh($\gamma$,n) reaction cross section as a function of the incident $\gamma$ energy. The dots denotes the folded cross section and the line with shaded area is the unfolded (monoenergetic) cross section. (b) Monoenergetic cross section.}
\end{figure}

The folded and monochromatic cross section of $^{103}\mathrm{Rh}(\gamma,\text{n})^{102}\mathrm{Rh}$, together with their uncertainties, are shown in Fig.~\ref{Folded}. The systematic uncertainty is \(3.15\%\), which originates from FED efficiency error ($3.02\%$), BGO gamma-spectrum error ($0.90\%$), and target thickness error ($0.10\%$). The methodological uncertainty is about $2.2\%$, resulting from the neutron count algorithm error ($2\%$) and the gamma-spectrum unfolding method error ($1\%$). According to Eq.~\eqref{sig_exp}, the statistical uncertainty is dominated by neutron count $N_{n}$ (3\%), and the statistical uncertainty of $N_{\gamma}$ is negligible because gamma count is sufficiently high. In this work, the total uncertainty comprises statistical, systematic, and methodological components. The total uncertainty of the unfolded cross section $\sigma(E_{\gamma})$ is less than $5\%$, except in the low cross section region near $S_{n}$ ($S_{n}<E_{\gamma}<10.0\ \mathrm{MeV}$) with $\sigma(E_{\gamma})<15\ \mathrm{mb}$. The monoenergetic cross section and their uncertainties are tabulated in  Table.~\ref{photoneutron}.
\setlength{\tabcolsep}{2.5pt}
\begin{table}[htbp]
\footnotesize
\centering
\caption{\label{photoneutron}Measured cross section and uncertainties for $^{103}$Rh($\gamma$,n)$^{102}$Rh}
\begin{tabular}{cccccc}
\toprule
\makecell{Energy \\ (MeV)} & 
\makecell{Cross \\ sections (mb)} & 
\makecell{Statistical \\ error (mb)} & 
\makecell{Methodological \\ error (mb)} & 
\makecell{Systematical \\ error (mb)} & 
\makecell{Total \\ error (mb)} \\
\midrule
9.57 & 6.35 & 0.56 & 0.11 & 0.19 & 0.87  \\ 
9.95 & 14.72 & 0.32 & 0.17 & 0.28 & 0.77  \\ 
10.33 & 19.13 & 0.21 & 0.25 & 0.42 & 0.88  \\ 
10.71 & 22.40 & 0.18 & 0.29 & 0.53 & 1.00  \\ 
11.09 & 26.52 & 0.21 & 0.35 & 0.65 & 1.22  \\ 
11.47 & 32.71 & 0.21 & 0.40 & 0.76 & 1.36  \\ 
11.85 & 41.56 & 0.29 & 0.48 & 0.90 & 1.66  \\ 
12.22 & 53.16 & 0.29 & 0.57 & 1.06 & 1.91  \\ 
12.60 & 67.19 & 0.31 & 0.70 & 1.37 & 2.38  \\ 
12.97 & 83.06 & 0.31 & 0.82 & 1.55 & 2.68  \\ 
13.34 & 99.98 & 0.31 & 0.99 & 1.95 & 3.25  \\ 
13.71 & 117.13 & 0.33 & 1.20 & 2.40 & 3.94  \\ 
14.07 & 133.63 & 0.35 & 1.46 & 2.91 & 4.72  \\ 
14.43 & 148.71 & 0.38 & 1.77 & 3.52 & 5.67  \\ 
14.79 & 161.68 & 0.40 & 2.09 & 4.13 & 6.61  \\ 
15.14 & 172.04 & 0.39 & 2.19 & 4.31 & 6.90  \\ 
15.48 & 179.41 & 0.38 & 2.35 & 4.65 & 7.38  \\ 
15.82 & 183.61 & 0.41 & 2.58 & 4.97 & 7.96  \\ 
16.16 & 184.63 & 0.40 & 2.71 & 5.16 & 8.28  \\ 
16.49 & 182.58 & 0.38 & 2.65 & 5.32 & 8.35  \\ 
16.81 & 177.75 & 0.38 & 2.71 & 5.45 & 8.55  \\ 
\bottomrule
\end{tabular}
\end{table}

\begin{figure}[b!]
\centering
\includegraphics[width=0.95\linewidth]{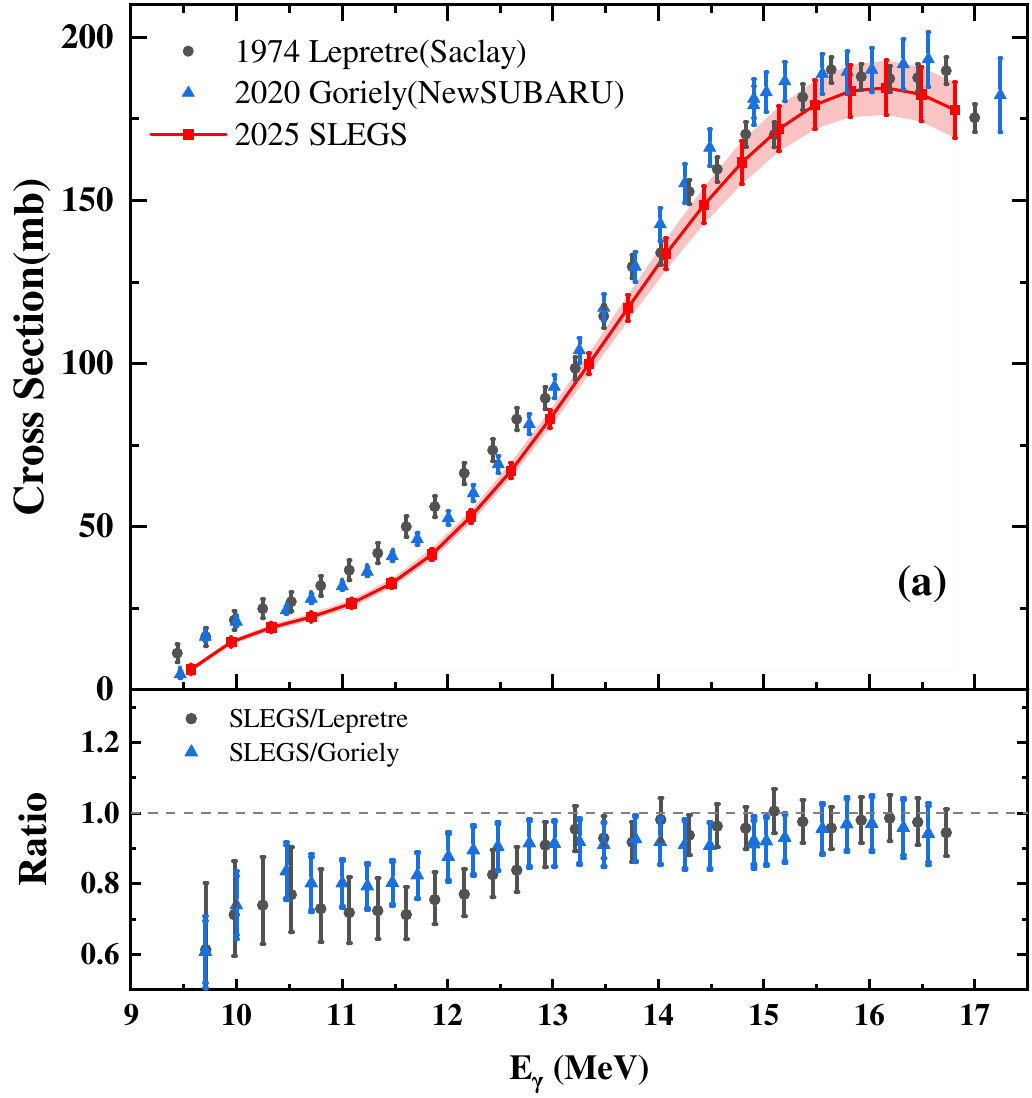}
\vspace{-0.5cm}
\includegraphics[width=0.95\linewidth]{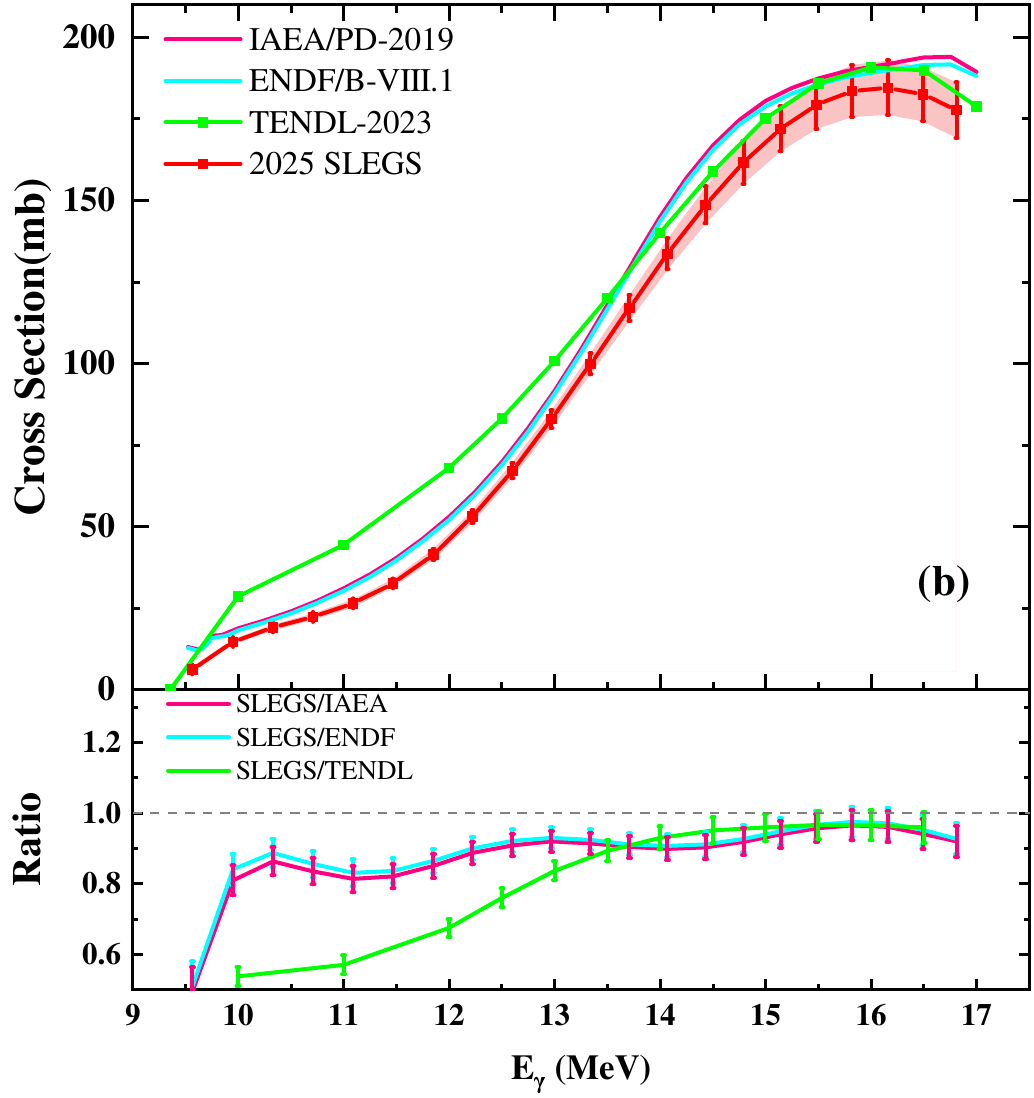}
\caption{\label{Mono}  (a) Comparison of cross section between SLEGS and (a) previous measurements, (b) evaluation libraries and their cross section ratios. 
}
\end{figure}

We compared the measured results with existing experimental and evaluated libraries and compared their ratios, as illustrated in Fig~\ref{Folded}(b) and Fig~\ref{Mono}. Overall, the present results show good agreement with the available data, albeit slightly lower. The ratio mostly lies in the range of 0.8--1.0, with a slight deviation observed near the $S_n$ region due to its relatively small cross section; nevertheless, the absolute discrepancy remains little. As mentioned in the introduction, overall, the $^{103}$Rh($\gamma$,n) cross section values obtained in the study are getting lower and lower, and this experiment measured an even lower cross section value, which may be due to the slant scattering mode of the SLEGS beamline, which can achieve more accurate energy sampling. In contrast, the cross section provided by TENDL-2023 exhibit higher values in the low-energy region. The outcomes of this study are slightly lower than those of previous studies, demonstrating good agreement with datasets of ENDF/B-$\mathrm{VIII}$.1, IAEA/PD-2019, and NewSUBARU~\cite{PhysRevC.102.064309} by LCS in $S_n$--13.5~MeV(both ENDF and IAEA assessments are based on NewSUBARU data). They show closer alignment with the results of Saclay~\cite{LEPRETRE197439} using the positron annihilation-in-flight technique and TENDL-2023 in 13.5~MeV--$S_{2n}$, although TENDL-2023 significantly overestimates the reaction cross section at $E_\gamma < 13.5$~MeV. 

To quantify the systematic differences between two cross section curves, their integral ratio over the $E_\gamma$ range from the neutron separation energy ($S_n$) to the two-neutron separation energy ($S_{2n}$) is considered~\cite{varlamov2019evaluation}. The total cross section within this energy interval is defined as:
\begin{equation}
    \sigma^\text{int} = \int_{S_n}^{S_{2n}} \sigma(E)\, dE.
\end{equation}
The $^{103}\text{Rh}(\gamma, n)^{102}\text{Rh}$ integrated cross section compared are $\sigma_{\text {SLEGS }}^{\text {int }} / \sigma_{\text {Lepetre }}^{\text {int }}=0.92 ,\quad \sigma_{\text {SLEGS }}^{\text {int }} / \sigma_{\text {Goriely }}^{\text {int }}=0.92,\quad \sigma_{\text {SLEGS }}^{\text {int }} / \sigma_{\text {ENDF }}^{\text {int }}=0.93,\quad \sigma_{\text {SLEGS }}^{\text {int }} / \sigma_{\text {IAEA }}^{\text {int }}=0.92$, and $\sigma_{\text {SLEGS }}^{\text {int }} / \sigma_{\text {TENDL }}^{\text {int }}=0.88$. This further proves the above discussion.

There have long been discrepancies between previous studies. The results of this study show consistency between the LCS $\gamma$-ray sources (NewSUBARU and SLEGS), which is crucial for improving nuclear data evaluation and optimizing theoretical model parameters.

\section{Summary} \label{sec:sum}
This study conducted high-precision measurements of the neutron capture and photoneutron reaction cross section of $^{103}\text{Rh}$, addressing long-standing data gaps and resolving historical experimental discrepancies. Using TOF techniques and $\rm{C_{6}D_{6}}$ detectors at CSNS Back-n, neutron capture cross section in the 1~eV--1~MeV energy range were systematically obtained. Several new resonance states were identified, their resonance parameters extracted, and MACS for $kT$ = 5--100~keV calculated. At the SSRF SLEGS platform, utilizing quasi-monoenergetic $\gamma$-ray beams and a total-efficiency detector array, combined with an innovative spectral unfolding method, monoenergetic photoneutron cross section of $^{103}\text{Rh}(\gamma$,n) were measured for the first time with an uncertainty below 5\%. The spectral features successfully reconciled historical conflicts between previous studies.

These results provide critical benchmarks for nuclear astrophysics (precise quantification of the s-process and p-process nucleosynthesis path), nuclear engineering (optimized design of self-sustaining neutron detectors), medical applications (isomeric state $^{103\mathrm{m}}\mathrm{Rh}$ potential for targeted therapy and radiological imaging), and nuclear data evaluations. Future work will improve the accuracy of neutron capture cross section and extend the measurement range of photoneutron reaction ($E_{\gamma}>S_{2n}$), deepen the theoretical investigation of newly observed resonance states, and promote the direct application of the data in stellar evolution models and nuclear technology systems.


\section*{Acknowledgements}
We thank the staff members of the Back-n white neutron facility (\href{https://cstr.cn/31113.02.CSNS.Back-n}{https://cstr.cn/31113.02.CSNS.Back-n}) at the CSNS (\href{https://cstr.cn/31113.02.CSNS}{https://cstr.cn/31113.02.CSNS}), for providing technical support and assistance in data collection and analysis. We also thank the SSRF of BL03SSID(\href{https://cstr.cn/31124.02.SSRF.BL03SSID}{https://cstr.cn/31124.02.SSRF.BL03SSID}) for the assistance on measurements and analysis.
This work was supported by the National Natural Science Foundation of China (Grant Nos. 12475152, 12147101, U1832182, and U2032137), and the Natural Science Foundation of Guangdong Province, China (Grant No. 2022A1515011184).

\bibliography{reference}

\providecommand{\noopsort}[1]{}\providecommand{\singleletter}[1]{#1}%
\begin{thebibliography}{10}
\expandafter\ifx\csname url\endcsname\relax
  \def\url#1{\texttt{#1}}\fi
\expandafter\ifx\csname urlprefix\endcsname\relax\def\urlprefix{URL }\fi
\expandafter\ifx\csname href\endcsname\relax
  \def\href#1#2{#2} \def\path#1{#1}\fi

\bibitem{meyer1994r}
B.~S. Meyer, The r-, s-, and p-processes in nucleosynthesis, Annu. Rev. Astron.
  Astr. 32 (1994).
\newblock \href {https://doi.org/10.1146/annurev.aa.32.090194.001101}
  {\path{doi:10.1146/annurev.aa.32.090194.001101}}.

\bibitem{gangopadhyay2019nuclear}
G.~Gangopadhyay, Nuclear inputs in astrophysical s-process nucleosynthesis, in:
  Proceedings of the DAE Symp. on Nucl. Phys, Vol.~64, 2019, p.~33,
  ~\url{https://www.sympnp.org/proceedings/64/I16.pdf}.

\bibitem{bao2000neutron}
Z.~Bao, H.~Beer, F.~K{\"a}ppeler, et~al., Neutron cross sections for
  nucleosynthesis studies, Atom. Data. Nucl. Data. 76 (2000) 70--154.
\newblock \href {https://doi.org/10.1006/adnd.2000.0838}
  {\path{doi:10.1006/adnd.2000.0838}}.

\bibitem{Reffo1982FastNC}
G.~Reffo, F.~Fabbri, K.~Wisshak, et~al., Fast neutron capture cross sections
  and related gamma-ray spectra of niobium-93, rhodium-103, and tantalum-181,
  Nucl. Sci. Eng. 80 (1982) 630--647.
\newblock \href {https://doi.org/10.1103/PhysRevLett.125.052001}
  {\path{doi:10.1103/PhysRevLett.125.052001}}.

\bibitem{XING201679}
J.~Xing, D.~Song, Y.~Wu, Hpr1000: Advanced pressurized water reactor with
  active and passive safety, Engineering 2 (2016) 79--87.
\newblock \href {https://doi.org/10.1016/J.ENG.2016.01.017}
  {\path{doi:10.1016/J.ENG.2016.01.017}}.

\bibitem{WU2024105110}
X.~Wu, J.~Jiang, T.~Wu, et~al., {Neutron sensitivity and uncertainty analysis
  of rhodium self-powered neutron detectors for reactor monitoring in HPR1000},
  Prog. Nucl. Energ. 170 (2024) 105110.
\newblock \href {https://doi.org/10.1016/j.pnucene.2024.105110}
  {\path{doi:10.1016/j.pnucene.2024.105110}}.

\bibitem{miller2011rhodium}
M.~E. Miller, M.~L. Sztejnberg, et~al., {Rhodium self-powered neutron detector
  as a suitable on-line thermal neutron flux monitor in BNCT treatments}, Med.
  Phys. 38 (2011) 6502--6512.
\newblock \href {https://doi.org/10.1118/1.3660204}
  {\path{doi:10.1118/1.3660204}}.

\bibitem{lee2003neutron}
S.~Lee, S.~Yamamoto, K.~Kobayashi, G.~Kim, J.~Chang, Neutron capture
  cross-section measurement of rhodium in the energy region from 0.003 ev to 80
  kev by linac time-of-flight method, Nuclear science and engineering 144~(1)
  (2003) 94--107.

\bibitem{popov1962energy}
Y.~P. Popov, F.~Shapiro, {Energy dependence of cross sections for (n, $\gamma$)
  reactions on a number of Odd-Z Nuclei}, J. Exp. Theor. Phys. 15
  (1962).~\url{http://jetp.ras.ru/cgi-bin/dn/e_015_04_0683.pdf}.

\bibitem{osti_4022848}
A.~D. Carlson, M.~P. Fricke, S.~J. Friesenhahn, {Rhodium neutron-capture cross
  section from 1 eV to 1 MeV}, Trans. Amer. Nucl. Soc. 14 (1971)
  352--353.~\url{https://www.osti.gov/biblio/4022848}.

\bibitem{le1976mesures}
C.~LeRigoleur, A.~Arnaud, J.~Taste, {Mesures en valeur absolue des sections
  efficaces de capture radiactivedes neutrons par le $^{23}$Na, Cr, $^{55}$Mn,
  Fe, Ni, $^{103}$Rh,Ta, $^{197}$Au, $^{238}$U dans le domaine de 10 {\`a} 600
  keV}, Report 4788 (1976). Centre d‘Etudes Nucleaires,
  Saclay.~\url{https://inis.iaea.org/records/1hgtc-dzh90}.

\bibitem{macklin1980100}
R.~Macklin, J.~Halperin, {$^{100,101,102,104}$Ru (n, $\gamma$) and $^{103}$Rh
  (n, $\gamma$) cross sections above 2.6 keV}, Nucl. Sci. Eng. 73 (1980)
  174--185.
\newblock \href {https://doi.org/10.13182/NSE80-A18697}
  {\path{doi:10.13182/NSE80-A18697}}.

\bibitem{Bokhovko}
M.~V. Bokhovko, V.~N. Kononov, E.~D. Poletaev, et~al., Average fast neutron
  radiative capture cross sections for fission products and for isotopes of
  rare earth elements, in: Nuclear Data for Science and Technology, 1992, pp.
  62--64.
\newblock \href {https://doi.org/10.1007/978-3-642-58113-7_16}
  {\path{doi:10.1007/978-3-642-58113-7_16}}.

\bibitem{shakilur2016measurement}
M.~Shakilur~Rahman, K.~Kim, G.~Kim, et~al., {Measurement of flux-weighted
  average cross-sections and isomeric yield ratios for $^{103}$Rh ($\gamma$,
  xn) reactions in the bremsstrahlung end-point energies of 55 and 60 MeV},
  Eur. Phys. J. A 52 (2016) 194.
\newblock \href {https://doi.org/10.1140/epja/i2016-16194-x}
  {\path{doi:10.1140/epja/i2016-16194-x}}.

\bibitem{KAWANO2020109}
T.~Kawano, Y.~Cho, P.~Dimitriou, et~al., Iaea photonuclear data library 2019,
  Nucl. Data Sheets 163 (2020) 109--162.
\newblock \href {https://doi.org/10.1016/j.nds.2019.12.002}
  {\path{doi:10.1016/j.nds.2019.12.002}}.

\bibitem{PAN2021109534}
W.~T. Pan, T.~Song, H.~Y. Lan, et~al., Photo-excitation production of medically
  interesting isomers using intense $\gamma$-ray source, Appl. Radiat. Isotopes
  168 (2021) 109534.
\newblock \href {https://doi.org/10.1016/j.apradiso.2020.109534}
  {\path{doi:10.1016/j.apradiso.2020.109534}}.

\bibitem{pang2023progress}
X.~Pang, B.~H. Sun, L.~H. Zhu, et~al., {Progress of photonuclear cross sections
  for medical radioisotope production at the SLEGS energy domain}, Nucl. Sci.
  Tech. 34 (2023) 187.
\newblock \href {https://doi.org/10.1007/s41365-023-01339-4}
  {\path{doi:10.1007/s41365-023-01339-4}}.

\bibitem{LEPRETRE197439}
A.~Leprêtre, H.~Beil, R.~Bergère, et~al., A study of the giant dipole
  resonance of vibrational nuclei in the 103 $\leq$ a $\leq$ 133 mass region,
  Nucl. Phys. A 219 (1974) 39--60.
\newblock \href {https://doi.org/10.1016/0375-9474(74)90081-5}
  {\path{doi:10.1016/0375-9474(74)90081-5}}.

\bibitem{varlamov2019evaluation}
V.~Varlamov, A.~Davydov, V.~Kaidarova, {Evaluation of reliable cross sections
  of photoneutron reactions on $^{103}$Rh and $^{165}$Ho nuclei}, Phys. Atom.
  Nucl. 82 (2019) 212--223.
\newblock \href {https://doi.org/10.1134/S1063778819030153}
  {\path{doi:10.1134/S1063778819030153}}.

\bibitem{PhysRevC.102.064309}
S.~Goriely, S.~P\'eru, G.~Col\`o, et~al., $\mathrm{E1}$ moments from a coherent
  set of measured photoneutron cross sections, Phys. Rev. C 102 (2020) 064309.
\newblock \href {https://doi.org/10.1103/PhysRevC.102.064309}
  {\path{doi:10.1103/PhysRevC.102.064309}}.

\bibitem{PhysRevResearch.6.013225}
G.~L. Yang, Z.~D. An, W.~Jiang, et~al., {Measurement of the neutron capture
  cross sections of rhenium up to stellar $s$- and $r$-process temperatures at
  the China Spallation Neutron Source Back-n facility}, Phys. Rev. Res. 6
  (2024) 013225.
\newblock \href {https://doi.org/10.1103/PhysRevResearch.6.013225}
  {\path{doi:10.1103/PhysRevResearch.6.013225}}.

\bibitem{tang2021back}
J.~Y. Tang, Q.~An, J.~B. Bai, et~al., {Back-n white neutron source at CSNS and
  its applications}, Nucl. Sci. and Tech. 32 (2021) 1--10.
\newblock \href {https://doi.org/10.1007/s41365-021-00846-6}
  {\path{doi:10.1007/s41365-021-00846-6}}.

\bibitem{ren2019c}
J.~Ren, X.~Ruan, J.~Bao, et~al., {The $\rm{C_{6}D_{6}}$ detector system on the
  Back-n beam line of CSNS}, Radiat. Detect. Technol. Methods 3 (2019) 1--9.
\newblock \href {https://doi.org/10.1007/s41605-019-0129-8}
  {\path{doi:10.1007/s41605-019-0129-8}}.

\bibitem{an2023measurement}
Z.~D. An, W.~W. Qiu, W.~Jiang, et~al., Measurement of the $^{181}$ta (n,
  $\gamma$) cross sections up to stellar s-process temperatures at the csns
  back-n, Sci. Rep. 13 (8 2023).
\newblock \href {https://doi.org/10.1038/s41598-023-39603-7}
  {\path{doi:10.1038/s41598-023-39603-7}}.

\bibitem{borella2007use}
A.~Borella, G.~Aerts, F.~Gunsing, et~al., The use of $\rm{C_{6}D_{6}}$
  detectors for neutron induced capture cross-section measurements in the
  resonance region, Nucl. Instrum. Meth. A 577 (2007) 626--640.
\newblock \href {https://doi.org/10.1016/j.nima.2007.03.034}
  {\path{doi:10.1016/j.nima.2007.03.034}}.

\bibitem{PhysRevC.108.035802}
X.~K. Li, Z.~D. An, W.~Jiang, et~al., {Measurement of the
  $^{141}\mathrm{Pr}$(n,$\ensuremath{\gamma})$ cross section up to stellar
  $s$-process temperatures at the China Spallation Neutron Source Back-n
  facility}, Phys. Rev. C 108 (2023) 035802.
\newblock \href {https://doi.org/10.1103/PhysRevC.108.035802}
  {\path{doi:10.1103/PhysRevC.108.035802}}.

\bibitem{Chen2019}
Y.~H. Chen, G.~Y. Luan, J.~Bao, et~al., {Neutron energy spectrum measurement of
  the Back-n white neutron source at CSNS}, Eur. Phys.J.A 55 (2019) 115.
\newblock \href {https://doi.org/10.1140/epja/i2019-12808-1}
  {\path{doi:10.1140/epja/i2019-12808-1}}.

\bibitem{Tain2002}
J.~L. Tain, F.~Gunsing, D.~aniel Cano, et~al., {Accuracy of the Pulse Height
  Weighting Technique for capture cross section measurements}, J. Nucl. Sci.
  Technol. 39 (2002) 689.
\newblock \href {https://doi.org/10.1080/00223131.2002.10875193}
  {\path{doi:10.1080/00223131.2002.10875193}}.

\bibitem{larson2008updated}
N.~M. Larson, {Updated user's guide for Sammy: Multilevel R-matrix fits to
  neutron data using Bayes' equations}, Tech. rep., Oak Ridge National
  Lab.(ORNL), Oak Ridge, TN (United States) (10 2008).
\newblock \href {https://doi.org/10.2172/941054} {\path{doi:10.2172/941054}}.

\bibitem{yang2023measurement}
G.~L. Yang, Z.~D. An, W.~Jiang, et~al., {Measurement of Br (n, $\gamma$) cross
  sections up to stellar s-process temperatures at the CSNS Back-n}, Nucl. Sci.
  Tech. 34 (2023) 1--16.
\newblock \href {https://doi.org/10.1007/s41365-023-01337-6}
  {\path{doi:10.1007/s41365-023-01337-6}}.

\bibitem{PhysRevC.100.045804}
A.~Gawlik, C.~Lederer-Woods, J.~Andrzejewski, et~al., {Measurement of the
  $^{70}\mathrm{Ge}(\mathrm{n},\ensuremath{\gamma})$ cross section up to 300
  keV at the CERN n\_TOF facility}, Phys. Rev. C 100 (2019) 045804.
\newblock \href {https://doi.org/10.1103/PhysRevC.100.045804}
  {\path{doi:10.1103/PhysRevC.100.045804}}.

\bibitem{brusegan2005neutron}
A.~Brusegan, E.~Berthoumieux, A.~Borella, et~al., Neutron capture and
  transmission measurements on $^{103}${Rh} down to thermal energies, in: AIP
  Conference Proceedings, Vol. 769, American Institute of Physics, 2005, pp.
  953--956.
\newblock \href {https://doi.org/10.1063/1.1945162}
  {\path{doi:10.1063/1.1945162}}.

\bibitem{Koning2023}
A.~Koning, S.~Hilaire, S.~Goriely, Talys: modeling of nuclear reactions, Eur.
  Phys. J. A 59 (2023) 131.
\newblock \href {https://doi.org/10.1140/epja/s10050-023-01034-3}
  {\path{doi:10.1140/epja/s10050-023-01034-3}}.

\bibitem{li2022measurement}
X.~K. Li, Z.~D. An, W.~Jiang, et~al., {Measurement of the Eu (n,$\gamma$) cross
  section up to 500 keV at the CSNS Back-n facility, and the stellar Eu
  (n,$\gamma$) cross section at s-process temperatures}, Eur. Phys. J. A 58 (12
  2022).
\newblock \href {https://doi.org/10.1140/epja/s10050-022-00887-4}
  {\path{doi:10.1140/epja/s10050-022-00887-4}}.

\bibitem{wang2022commissioning}
H.~W. Wang, G.~T. Fan, L.~X. Liu, et~al., {Commissioning of laser electron
  gamma beamline SLEGS at SSRF}, Nucl. Sci. Tech. 33 (2022) 87.
\newblock \href {https://doi.org/10.1007/s41365-022-01076-0}
  {\path{doi:10.1007/s41365-022-01076-0}}.

\bibitem{7370951}
H.~Xu, G.~Fan, H.~Wu, et~al., {Interaction chamber design for an energy
  continuously tunable sub-Mev Laser-Compton $\gamma$-ray source}, IEEE
  Transactions on Nuclear Science 63 (2016) 906--912.
\newblock \href {https://doi.org/10.1109/TNS.2015.2496256}
  {\path{doi:10.1109/TNS.2015.2496256}}.

\bibitem{XU2025170249}
H.~Xu, H.~Utsunomiya, G.~Fan, et~al., {Gamma-ray flux in Gated CW operation of
  $\mathrm{CO_2}$ laser at SLEGS}, Nucl. Instrum. Meth. A 1073 (2025) 170249.
\newblock \href {https://doi.org/10.1016/j.nima.2025.170249}
  {\path{doi:10.1016/j.nima.2025.170249}}.

\bibitem{li2025new}
Z.~C. Li, Z.~R. Hao, Q.~K. Sun, et~al., {New measurement of $^{63}$Cu
  ($\gamma$, n) $^{62}$Cu cross-section using quasi-monoenergetic $\gamma$-ray
  beam}, Nucl. Sci. Tech. 36 (2025) 34.
\newblock \href {https://doi.org/10.1007/s41365-024-01631-x}
  {\path{doi:10.1007/s41365-024-01631-x}}.

\bibitem{LIU2024169314}
L.~X. Liu, H.~Utsunomiya, G.~T. Fan, et~al., {Energy profile of laser Compton
  slant-scattering $\gamma$-ray beams determined by direct unfolding of
  total-energy responses of a BGO detector}, Nucl. Instrum. Meth. A 1063 (2024)
  169314.
\newblock \href {https://doi.org/10.1016/j.nima.2024.169314}
  {\path{doi:10.1016/j.nima.2024.169314}}.

\bibitem{HAO20204pi}
Z.~R. Hao, G.~T. Fan, L.~X. Liu, et~al., {Design and simulation of 4$\pi$
  flat-efficiency $^{3}$He neutron detector array}, Nucl. Tech. 43 (2020)
  57--65.
\newblock \href {https://doi.org/10.11889/j.0253-3219.2020.hjs.43.110501}
  {\path{doi:10.11889/j.0253-3219.2020.hjs.43.110501}}.

\bibitem{jiao2025measurements}
P.~Jiao, Z.~R. Hao, Q.~K. Sun, et~al., {Measurements of $^{27}$Al ($\gamma$, n)
  reaction using quasi-monoenergetic $\gamma$ beams from 13.2 to 21.7 MeV at
  SLEGS}, Nucl. Sci. Tech. 36 (2025) 66.
\newblock \href {https://doi.org/10.1007/s41365-025-01662-y}
  {\path{doi:10.1007/s41365-025-01662-y}}.

\bibitem{hao2025photoneutron}
Z.~R. Hao, L.~X. Liu, Y.~Zhang, et~al., {Photoneutron cross-section data
  generation and analysis at the Shanghai laser electron gamma source}, Nucl.
  Sci. Tech. 36 (2025) 183.
\newblock \href {https://doi.org/10.1007/s41365-025-01773-6}
  {\path{doi:10.1007/s41365-025-01773-6}}.

\bibitem{PhysRev.162.1098}
B.~L. Berman, J.~T. Caldwell, R.~R. Harvey, et~al., {Photoneutron Cross
  Sections for $^{90}{\mathrm{Zr}}$, $^{91}{\mathrm{Zr}}$,
  $^{92}{\mathrm{Zr}}$, $^{94}{\mathrm{Zr}}$, and $^{89}{\mathrm{Y}}$}, Phys.
  Rev. 162 (1967) 1098--1111.
\newblock \href {https://doi.org/10.1103/PhysRev.162.1098}
  {\path{doi:10.1103/PhysRev.162.1098}}.

\bibitem{RevModPhys.47.713}
B.~L. Berman, S.~C. Fultz, Measurements of the giant dipole resonance with
  monoenergetic photons, Rev. Mod. Phys. 47 (1975) 713--761.
\newblock \href {https://doi.org/10.1103/RevModPhys.47.713}
  {\path{doi:10.1103/RevModPhys.47.713}}.

\bibitem{HAO2025}
Z.~R. Hao, G.~T. Fan, H.~W. Wang, et~al., {The day-one experiment at SLEGS:
  systematic measurement of the ($\gamma$, 1n) cross sections on $^{197}$Au and
  $^{159}$Tb for resolving existing data discrepancies}, Sci. Bull. (2025).
\newblock \href {https://doi.org/10.1016/j.scib.2025.05.037}
  {\path{doi:10.1016/j.scib.2025.05.037}}.

\bibitem{PhysRevC.19.1684}
J.~W. Jury, B.~L. Berman, D.~D. Faul, et~al., Photoneutron cross sections for
  $^{13}\mathrm{C}$, Phys. Rev. C 19 (1979) 1684--1692.
\newblock \href {https://doi.org/10.1103/PhysRevC.19.1684}
  {\path{doi:10.1103/PhysRevC.19.1684}}.

\bibitem{PhysRevC.90.064616}
D.~M. Filipescu, I.~Gheorghe, H.~Utsunomiya, et~al., Photoneutron cross
  sections for samarium isotopes: Toward a unified understanding of
  ($\ensuremath{\gamma}$,n) and (n,$\ensuremath{\gamma}$) reactions in the rare
  earth region, Phys. Rev. C 90 (2014) 064616.
\newblock \href {https://doi.org/10.1103/PhysRevC.90.064616}
  {\path{doi:10.1103/PhysRevC.90.064616}}.

\bibitem{GUTTORMSEN1996371}
M.~Guttormsen, T.~Tveter, L.~Bergholt, et~al., The unfolding of continuum
  $\gamma$-ray spectra, Nucl. Instrum. Meth. A 374 (1996) 371--376.
\newblock \href {https://doi.org/10.1016/0168-9002(96)00197-0}
  {\path{doi:10.1016/0168-9002(96)00197-0}}.

\bibitem{PhysRevC.109.014617}
H.~Utsunomiya, S.~Goriely, M.~Kimura, et~al., Photoneutron emission cross
  sections for $^{13}\mathrm{C}$, Phys. Rev. C 109 (2024) 014617.
\newblock \href {https://doi.org/10.1103/PhysRevC.109.014617}
  {\path{doi:10.1103/PhysRevC.109.014617}}.

\bibitem{PhysRevC.98.054619}
H.~Utsunomiya, T.~Renstr\o{}m, G.~M. Tveten, et~al., {Photoneutron cross
  sections for Ni isotopes: Toward understanding (n,$\ensuremath{\gamma}$)
  cross sections relevant to weak $s$-process nucleosynthesis}, Phys. Rev. C 98
  (2018) 054619.
\newblock \href {https://doi.org/10.1103/PhysRevC.98.054619}
  {\path{doi:10.1103/PhysRevC.98.054619}}.

\end{thebibliography}
\end{document}